\documentclass[aps,physrev,twocolumn,superscriptaddress]{revtex4-2}

\usepackage{siunitx}
\usepackage{graphicx}
\usepackage{amsmath}
\usepackage{longtable}
\usepackage{xcolor}

\begin{document}

\title{Broad-band ellipsometry study of the anisotropic dielectric response of YAlO$_\text{3}$}

\author{Laurent Bugnon}
\affiliation{Department of Physics, University of Fribourg, CH-1700 Fribourg, Switzerland}
\author{Yurii G. Pashkevich}
\affiliation{Department of Physics, University of Fribourg, CH-1700 Fribourg, Switzerland}
\affiliation{Galkin Donetsk Institute for Physics and Engineering of NASU, 03028 Kyiv, Ukraine}
\author{Christian Bernhard}
\author{Premysl Marsik}
\affiliation{Department of Physics, University of Fribourg, CH-1700 Fribourg, Switzerland}

\date{\today}

\begin{abstract}
We present a broad band (THz to UV) ellipsometry study of the anisotropic dielectric response of the orthorhombic perovskite YAlO$_\text{3}$. The ellipsometric measurements have been performed on YAlO$_\text{3}$ crystals with three different surface cuts and for six high symmetry configurations of the crystal axes with respect to the plane of incidence of the photons. The obtained data are presented in terms of the Mueller Matrix elements $N$, $C$, and $S$ and their features are analyzed and discussed with respect to the anisotropy of the dielectric response tensor. In particular, in the infrared range we have identified all 25 infrared active phonon modes that have been predicted from theoretical studies. We also discuss a negative refraction effect that naturally occurs in the vicinity of an anisotropic longitudinal-optical phonon. Moreover, we have determined the temperature dependence of the phonon parameters between 10 and \SI{330}{\kelvin}. The dielectric response above the phonon range, from about \SIrange{0.1}{6.5}{\electronvolt}, is shown to be featureless and characteristic of an insulator with a large band gap above \SI{6.5}{\electronvolt} and is well described by anisotropic Cauchy model. 
\end{abstract}

\maketitle

\section{Introduction}
Yttrium Aluminate (YAlO$_\text{3}$, also known as YAP – Yttrium Aluminum Perovskite) can be readily grown as large high quality single crystals. It is transparent in the visible range with a moderate refractive index (of about 1.93 at \SI{2}{\electronvolt}) and a small birefringence. The phonon spectrum in the far-infrared range is fairly rich and exhibits a clear anisotropy. YAlO$_\text{3}$ is also a hard material with a rather high thermal conductivity. For some of those qualities, it has been proposed as a material for hosting lasing dopants \cite{Bagdasarov1969,Weber1969,Kaminskii2014}, as alternative for the well-known yttrium-aluminum-garnet Y$_\text{3}$Al$_\text{5}$O$_\text{12}$ (YAG).

YAlO$_\text{3}$ is also frequently used as a substrate for the epitaxial growth of thin films from unconventional oxides. In particular, the lattice parameters of YAlO$_\text{3}$ are well matched with those of several interesting perovskite manganites and their anisotropy can be utilized for the tailored growth of strain-engineered samples \cite{Shimamoto2017,Vistoli2018}. The aim of the present work is to lay the ground for the optical spectroscopy of such thin films and heterostructures which require a detailed a priori knowledge of the anisotropic response of the YAlO$_\text{3}$ substrate.

The Raman response of YAlO$_\text{3}$ has been already studied in great detail \cite{Salje1974,Suda2003,Chopelas2011,HernandezRodriguez2017,Dewo2020}. Theoretical studies of the IR and Raman phonon spectra of YAlO$_\text{3}$ have also been reported \cite{Gupta1999,Suda2003,Vali2007,Dewo2020}. However, the corresponding detailed optical studies of the IR dielectric tensor and the anisotropy of the phonon response are still missing.

Infrared spectroscopic ellipsometry has proven to be an excellent technique for studies of the phonon response of ionic crystals \cite{Schubert2005}, especially of those with a strong optical anisotropy. For isotropic bulk materials, ellipsometry allows one to directly measure the complex dielectric function. Unlike IR reflectivity measurements which probe only the amplitude of the reflected signal (but not the phase), it does not require a Kramers-Kronig transformation and the related extrapolation of the data to zero and infinite frequency.

For anisotropic crystals, the interpretation of the ellipsometry spectra is more complicated since the probing light interacts with the sample under general angle of incidence and polarization state. In general, the measured spectra can be affected by all the elements of the dielectric tensor. In such cases, several ellipsometry measurements must be performed on different surfaces and for various orientations of the crystal axes with respect to the plane of incidence and the combined data need to be analyzed with an anisotropic model that allows for finite diagonal and off-diagonal elements of the dielectric tensor. For materials with tetragonal or orthorhombic symmetry, this can be done on high symmetry cuts \cite{Humlicek2000,Schubert2000,Schoeche2013,Tumenas2017} or with a sequence of different orientations of single low-symmetry cut crystal \cite{Mock2019}.

In the following, we present a broad-band (THz to UV) ellipsometry study of the dielectric tensor of orthorhombic YAlO$_\text{3}$. The measurements are performed on crystals with high symmetry surface cuts and for high symmetry orientations of the crystal axes with respect to the plane of incidence of the photon beam. We present a detailed analysis of ellipsometric data of all 6 high symmetry configurations and document the temperature dependence of the YAlO$_\text{3}$ phonon parameters.

As stated above, the intention of this study is to allow for subsequent ellipsometric analysis of the optical response of thin films deposited on top of the YAlO$_\text{3}$ substrates. For that reason we focus on interpretation of the features observed in raw ellipsometric spectra in the form of NCS Mueller matrix elements.

\begin{figure}
\includegraphics[width=0.45\textwidth]{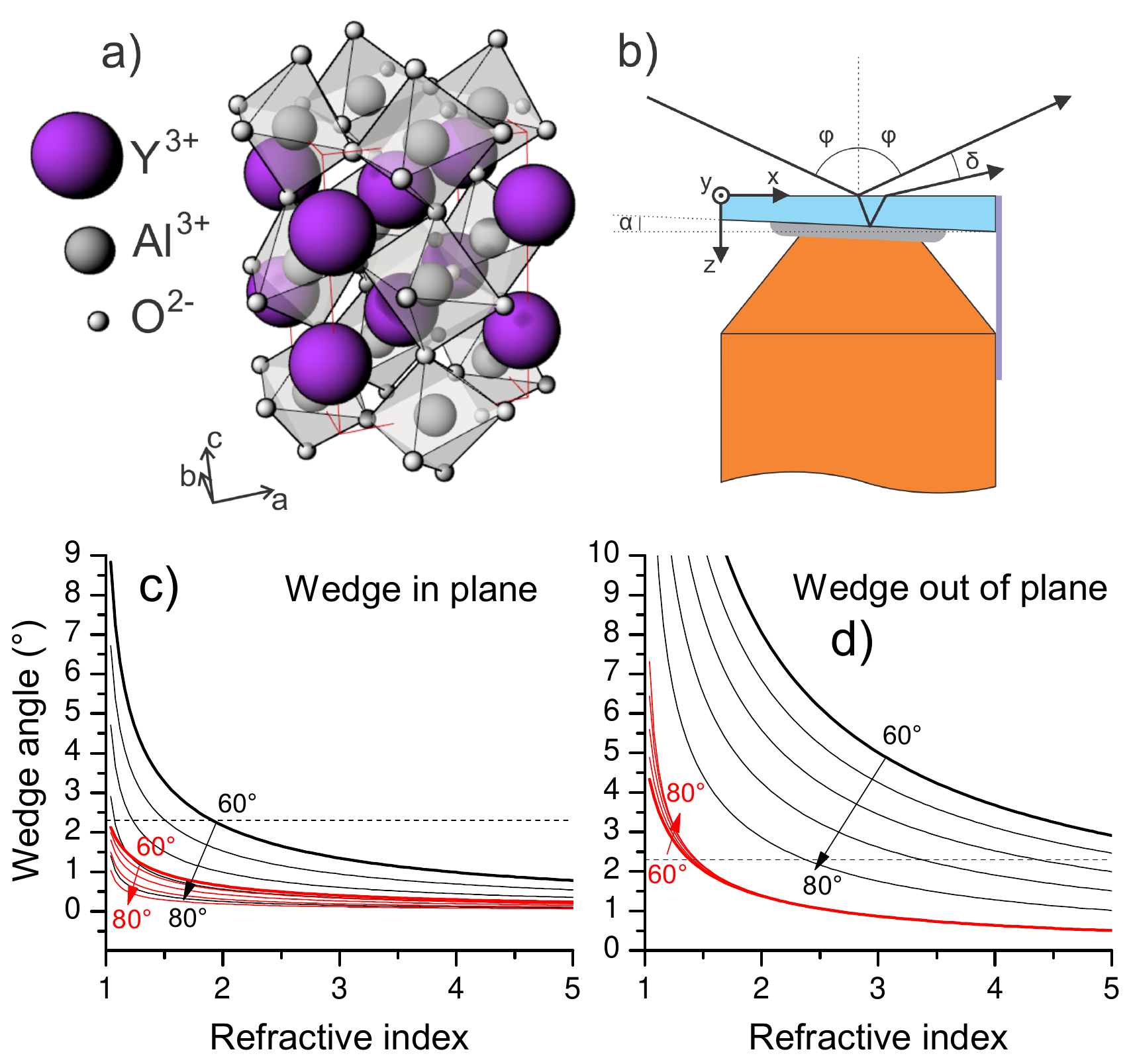}
\caption{\label{YAO_schema} a) Sketch of the distorted perovskite structure of YAlO$_\text{3}$ with the unit cell indicated by the red lines and the atomic position adapted from \cite{Suda2003}. b) Sketch of a YAlO$_\text{3}$ crystal (light blue) with a wedge angle, $\alpha$, that is mounted with silver paint (gray) on a copper sample holder (orange). Also shown is the coordinate system of the ellipsometer with the plane of incidence of the photon beam defined by the $x$- and $z$-axes and the incident and reflected beam at the angle $\varphi$ with respect to the surface normal. The beam reflected from the backside is deviated by the angle $\delta$. The side of the sample is covered by aluminum tape (violet) to further reduce scattered and unwanted light. c) Wedge angle along the plane of incidence that is required for a deviation of $\delta = \SI{5}{\degree}$ (red lines) or for a total internal reflection (black lines) calculated as a function of the refractive index and the incidence angle $\varphi$ ranging from \SI{60}{\degree} (thick lines) to \SI{80}{\degree} (indicated by the arrows). d) Same as (c) but with the crystal rotated by \SI{90}{\degree}, i.e. with the wedge perpendicular to the plane of incidence. Horizontal dashed lines mark a wedge angle of \SI{2.3}{\degree}.}
\end{figure}

\section{Samples and Experiment Setups}
YAlO$_\text{3}$ crystallizes in a distorted perovskite structure with orthorhombic symmetry as displayed in Fig. \ref{YAO_schema}a. Using the $Pbnm$ notation, the orthorhombic lattice parameters are $a=\SI{5.18}{\angstrom}$, $b=\SI{5.33}{\angstrom}$ and $c=\SI{7.375}{\angstrom}$ \cite{Diehl1975,SurfaceNetData}. Throughout the manuscript, we will use the notation $a$-, $b$-, and $c$-axis for the crystal axes [100], [010], [001], respectively, and $a$-cut, $b$-cut and $c$-cut for the (100), (010), (001) surface cuts, respectively. The coordinate system $x$, $y$, $z$ will be used to describe the ellipsometry setup and the orientation of the crystal surface with respect to the plane of incidence of the incoming and reflected phonon beam, as shown in Fig. \ref{YAO_schema}b.

The ellipsometry experiments have been performed on a set of YAlO$_\text{3}$ crystals with high symmetry surface cuts along (100), (010), and (001) that have been purchased from a commercial vendor (SurfaceNet). For the FIR to UV ranges we used crystals with dimensions of $5\times 5 \times \SI{0.5}{\cubic\milli\metre}$, in the THz range we used two additional, larger crystals with $10 \times 10 \times \SI{0.5}{\cubic\milli\metre}$.

The $5\times 5 \times\SI{0.5}{\cubic\milli\metre}$ samples have been wedged by mechanical polishing of the backside surface such that it is inclined by an angle of $\alpha = \SI{2.3}{\degree}$ with respect to front surface (see Fig. \ref{YAO_schema}b). The wedge has been oriented for the $a$-cut sample in the $ab$-plane; for the $b$-cut sample in $bc$-plane; and for the $c$-cut sample in the $ca$-plane. This wedging causes a deviation between the primary beam that is reflected from the front surface and the one reflected from the back surface by an angle of $\delta$ (indicated in Fig. \ref{YAO_schema}b). This ensures that the light reflected from the back surface does not reach the detector, in the frequency ranges for which YAlO$_\text{3}$ is fully transparent. Moreover, for a large enough angle of incidence of the photons, $\varphi$, the beam reflected from the backside cannot even exit the sample, due to the total internal reflection at the front surface. Fig. \ref{YAO_schema}c shows the critical wedge angle for this total internal reflection effect (black lines) and for a deviation of the beam by $\delta = \SI{5}{\degree}$ (red lines). The latter value corresponds to the maximal acceptance angle of the focusing mirrors of the infrared ellipsometer.

Note that the wedging also helps to avoid unwanted backside reflection if the sample is rotated by \SI{90}{\degree} around the $z$-axis – which is the case when measuring with the plane of incidence along the other high symmetry axis of the crystal. As shown in Fig. \ref{YAO_schema}d, a much larger wedging angle is required here to completely suppress the backside-reflected beam due to the total internal reflection. This is because the wedge angle lies in a plane perpendicular to the plane of incidence and therefore does not add up to the incidence angle. Nevertheless, as long as the refractive index is large enough, i.e. for $n\geq1.5$, the wedge angle of \SI{2.3}{\degree} leads to deviation of $\delta > \SI{5}{\degree}$ out of the plane of incidence that ensures that the beam from the backside does not reach the detector.

The ellipsometry measurements in the infrared range from about \SIrange{40}{5000}{\per\centi\meter} have been performed with a home-built setup that is attached to a Bruker 70v FTIR spectrometer and has the infrared beam-path in a rough vacuum of about $\SI{e-1}{\milli\bar}$. A more detailed description of the ellipsometer can be found in \cite{Bernhard2004}. It is equipped with a He-flow cryostat (CryoVac) for measurements at temperatures from \SIrange{10}{350}{\kelvin}.

In the far-infrared range from \SIrange{40}{700}{\per\centi\meter}, we used a Hg arc lamp, a solid Si beam-splitter, and a \SI{1.2}{\kelvin} bolometer (Infrared Laboratories) as detector. The ellipsometric measurements have been performed in rotating analyzer mode (RAE) with a fixed polarizer (P) in front of the sample (S) and a rotating analyzer (A$_\text{R}$) after the sample (P-S-A$_\text{R}$ configuration). The polarizers made from a metal wire grid on polyethylene substrate have been purchased from a commercial vendor (Tydex). For some measurements, we also used an optional, static compensator (C) that is based on the total internal reflection in a prism of made from undoped Si (P-C-S-A$_\text{R}$ configuration). Since the raw RAE data can be affected by imperfections of the polarizers, and a polarization sensitivity of the detector, the measured data have been treated with a correction algorithm similar to that described in \citep{Boer1993}. The details will be discussed elsewhere, together with the calibration procedure for the offsets of the polarizers and the retarder.

In the mid-infrared range, from about \SIrange{500}{5000}{\per\centi\meter}, we used a Globar source, a Germanium coated KBr beam-splitter, and a \SI{4.2}{\kelvin} bolometer (Infrared Laboratories) as detector. For the latter, the Winston cone has been replaced with a cold focusing optics as to reduce the polarization sensitivity. The polarizer in front of the sample consisted of a high-quality free standing gold wire grid (InfraSpecs). The polarization state after the reflection from the sample has been analyzed with a rotating compensator (C$_\text{R}$) followed by a static analyzer (P-S-C$_\text{R}$-A configuration). The rotating compensator is based on the internal reflection in a ZnSe prism that is mounted together with three gold mirrors in a four-bounce configuration that can be rotated without changing the beam path to the detector, similar as in \cite{Stanislavchuk2013}. The static analyzer (typically at A=0) consisted of a tandem of KRS5 based grid polarizers (Specac).

All the far- and mid-infrared data have been collected with a resolution of \SI{2}{\per\centi\meter} and at an incidence angle of $\varphi = \SI{75}{\degree}$. The mid-infrared RCE and far-infrared RAE spectra have shown a very good overlap such that we could simply merge the data, either at \SI{585}{\per\centi\meter} or at \SI{540}{\per\centi\meter}. The upper frequency limit of the far-infrared data was determined by the multi-phonon absorption band of Si (beam-splitter and the optional static compensator) between 600-\SI{620}{\per\centi\meter}. The lower limit of the MIR data was determined by the phonon absorption in the ZnSe prism which sets in below \SI{520}{\per\centi\meter}.

The high energy spectra from \SIrange{0.5}{6.5}{\electronvolt} (\SIrange{4033}{52423}{\per\centi\meter}), spanning the near-infrared, visible, and UV ranges, have been measured with an IR-extended J. A. Woollam VASE instrument, at four angles of incidence (55, 65, 75 and \SI{85}{\degree}), with a spectral resolution of \SI{0.05}{\electronvolt}, in an ambient atmosphere.

The low energy spectra in the THz range from \SIrange{0.1}{2.5}{\tera\hertz} (\SIrange{3}{83}{\per\centi\meter}) have been obtained with a home-built time-domain THz ellipsometer, as described in \cite{Marsik2016}. This instrument has been operated in a rotating-analyzer (RAE) mode at an incidence angle of $\varphi = \SI{75}{\degree}$. The phase sensitive time-domain detection technique enables one to determine the ellipsometric angles $\Psi$ and $\Delta$, the latter in the full \SI{360}{\degree} range without the need of additional compensator. The time-resolved detection scheme also allows one to distinguish between the signals that arise from the reflections on the front surface and the back-side that are delayed in time. Accordingly, the contribution of the latter back-side reflection can be removed by cutting the data above a certain time delay. We have used a time window of \SI{15.5}{\pico\second} which results in a spectral resolution of about \SI{1.5}{\per\centi\meter}. To minimize artifacts that may arise from diffraction effects that become increasingly important towards longer wavelength \cite{Humlicek2004}, we have used the larger 10x10x\SI{0.5}{\cubic\milli\metre} crystals for the THz ellipsometry measurements.

Corresponding broad-band (THz to UV) spectroscopic ellipsometry experiments using the above described combination of ellipsometers have been previously reported for various thin films on isotropic substrates \cite{Mallett2016,Marsik2016,Sen2017,Marsik2022} and for anisotropic samples \cite{Roessle2013,YazdiRizi2017}.

\section{Data analysis}
The orthorhombic unit cell with $Pbnm$ space group ($D^{16}_{2h}$) of YAlO$_\text{3}$ (see Fig. \ref{YAO_schema}a) contains 4 formula units with 20 atoms and in total 60 degrees of freedom. From all the possible displacement modes, besides the acoustic, Raman-active or silent ones, 25 modes are reported to be infrared-active \cite{Gupta1999,Vali2007}. The orthorhombic symmetry yields a diagonal dielectric tensor $\varepsilon$ with the principal axes oriented along the $a$, $b$ and $c$ lattice directions and 9 $B_{1u}$ modes in the $\varepsilon_a$ component, 9 $B_{3u}$ modes in $\varepsilon_b$, and 7 $B_{2u}$ modes in $\varepsilon_c$.

The samples were measured in high symmetry configurations for which the $x$, $y$, and $z$ axes, that define the plane of incidence of the ellipsometer and the measured surface of the YAlO$_\text{3}$ crystal (see Fig. \ref{YAO_schema}b), coincide with the $a$, $b$ and $c$ axes directions of the YAlO$_\text{3}$ crystals for any of the 6 possible permutations. For these high symmetry configurations the birefringence does not give rise to a mixing of the $p$- and $s$- components. In the following, we describe the reflection and refraction in the framework of the $x$- and $z$-components of the wavevectors. For the sake of brevity, we use the geometrical components, $\boldsymbol{\kappa}$, that are related to the physical wavevector, $\boldsymbol{k}$, according to $\boldsymbol{k} = \boldsymbol{\kappa}\omega/c$. The incident wavevector (in vacuum) is denoted as $\boldsymbol{\kappa}_{inc} = (\xi, 0, q)$, with $\xi = \sin \varphi$ and $q = \cos\varphi$. The $x$-component $\xi$ is conserved for the reflected as well as the refracted beams in $p$- and $s$-polarization. Accordingly, the reflection problem can be treated as stationary in $x$ and independent of the $y$-component \cite{Berreman1972}. For the $z$-components of the two refracted beams denoted as $\kappa_p$ and $\kappa_s$, it follows that:		
\begin{equation}
\kappa_p = \sqrt{\varepsilon_x \left(1-\frac{\xi^2}{\varepsilon_z}\right)}, \quad \kappa_s = \sqrt{\varepsilon_y - \xi^2}, 
\label{kappa p s}
\end{equation}
where $\varepsilon_x$, $\varepsilon_y$ and $\varepsilon_z$ are the elements of the diagonal dielectric tensor, i.e. one of the permutations of $\varepsilon_a$, $\varepsilon_b$ and $\varepsilon_c$. The Jones matrix thus acquires a diagonal shape, with the following Fresnel coefficients $r_p$ and $r_s$ on the diagonal: 
\begin{equation}
r_p = \frac{\kappa_p - \varepsilon_x q}{\kappa_p + \varepsilon_x q}, \quad r_s = \frac{q - \kappa_s}{q + \kappa_s}.
\label{r p s}
\end{equation}
Note that we employ the Fresnel’s convention for $r_p$ \cite{Tompkins2004}.

The ellipsometric response is traditionally written in terms of the complex ratio $\rho = r_p/r_s = \tan \Psi \exp(i\Delta)$, with the ellipsometric angles $\Psi$ and $\Delta$. In RAE or RCE mode, the directly accessible quantities are the components of the normalized Muller matrix, which in the high symmetry configuration with diagonal Jones matrices acquires the following block-diagonal form,
\begin{equation}
\hat{M} = \begin{bmatrix}
1 & N & 0 & 0\\
N & 1 & 0 & 0\\
0 & 0 & C & S\\
0 & 0 & -S & C
\end{bmatrix}.
\label{NCS matrix}
\end{equation}
The elements of this so-called NCS-form of the Mueller matrix are related to the ellipsometric angles by $N=-\cos2\Psi$, $C=\sin2\Psi \cos\Delta$, and $S = \sin2\Psi \sin\Delta$. Effects from depolarization are not considered here.

In the simplest case of RAE in P-S-A$_\text{R}$ configuration, with the polarizer set to $P=\SI{45}{\degree}$, the modulation of the intensity at the detector, as a function of the analyzer azimuth A, is given by $I(A) = I_0 (1+ N \cos2A + C \sin2A)$, i.e. the $N$ and $C$ components are measured directly. In P-C-S-A$_\text{R}$ configuration, with an ideal $\lambda /4$-retarder as a static compensator and $P = \SI{45}{\degree}$, the intensity at the detector follows $I(A)=I_0(1+N\cos2A+S\sin2A)$, hence the remaining element $S$ of the sample NCS Mueller matrix is determined. The far-infrared data presented below have been obtained using this approach.

With a rotating compensator (RCE), in P-S-C$_\text{R}$-A configuration, as it has been used in the mid-infrared range, the modulation at the detector is
\small
\begin{align*}
I(B) =& I_0(1+\frac{N}{2+N}\cos4B + \frac{C}{2+N}\sin4B - \frac{S}{2+N}\sin2B),
\end{align*}
\normalsize
where $B$ is the azimuth of the rotating compensator. Therefore, RCE provides access to all three $N$, $C$, $S$ elements, with $P = \SI{45}{\degree}$ and $A = 0$, see e.g. \cite{Fujiwara2007}.

The detection of amplitude and phase by the time-domain THz spectroscopic technique allows measuring directly the complex ratio $\rho = \tan \Psi \exp i\Delta$, \cite{Nagashima2001,Neshat2012}, or, in a general sense, the components of the complex Jones matrix. To facilitate the comparison with the infrared data, the measured spectra in the THz range are also shown in terms of the Mueller matrix elements $N$, $C$, $S$.

The three diagonal elements of dielectric tensor of YAlO$_\text{3}$, $\varepsilon_a$, $\varepsilon_b$ and $\varepsilon_c$, are calculated in the infrared range with the following  factorized four-parameter model, as introduced by \cite{Berreman1968, Gervais1974} and discussed in the context of ellipsometry by Schubert et al. \cite{Schubert2000} and references therein:  
\begin{equation}
\varepsilon_j(\omega)=\varepsilon_{\infty j} \prod_{i=1}^{N_j} \frac{\omega^2_{\mathrm{LO}ji} - \omega^2 - i\omega\gamma_{\mathrm{LO}ji}}{\omega^2_{\mathrm{TO}ji} - \omega^2 - i\omega\gamma_{\mathrm{TO}ji}},
\label{FPSQ diel func}
\end{equation}
where $j = a, b, c$; $\varepsilon_{\infty j}$ is the high-frequency dielectric constant, $N_j = 9$ for $j=a,b$ and $N_j = 7$ for $j=c$. The parameters $\omega_{\text{TO}ji}$ and $\gamma_{\mathrm{TO}ji}$ account for the resonance frequency and the broadening of the $i$-th transversal optical (TO) phonon. The corresponding longitudinal optical (LO) phonons are characterized with frequencies, $\omega_{\text{LO}ji}$, and broadenings $\gamma_{\mathrm{LO}ji}$. 
From this dielectric tensor, the wavevector z-components $\kappa_p$ and $\kappa_s$ (Eq.\eqref{kappa p s}) have been calculated, for a given angle of incidence, $\varphi$, and the particular high symmetry orientation. Subsequently, the Fresnel coefficients (Eq.\eqref{r p s}), their ratio $\rho$, and finally the Mueller matrix elements have been derived and fitted to the experimental NCS data, by varying the parameters of the phonon oscillator parameters in Eq. \eqref{FPSQ diel func}.

As a curiosity, we also discuss how the negative refraction effects, which can occur in the vicinity of the longitudinal optical phonon modes ($\omega_\text{LO}$) of such anisotropic crystals, show up in the ellipsometric spectra. According to the original work of Veselago \cite{Veselago1968}, a negative refractive index can occur if both the real parts of the permittivity $\varepsilon$ and the permeability $\mu$ are negative. Of the two solutions of the complex square-root, $N=n+ik=\sqrt{\varepsilon\mu}$, the physically relevant one requires $k>0$ and thus $n<0$, such that the plane waves propagate in a direction that is opposite to the one of the energy flow (or the Pointing vector). This is equivalent to the statement that whenever the product of generally complex values of $\varepsilon$ and $\mu$ acquires a negative imaginary part, the physically correct solution of the square root is the one with positive imaginary part ($k>0$), eventually leading to $n<0$ \cite{Boardman2005}.

However, similar phenomena appear even in the absence of magnetic permeability resonances, and are caused by the anisotropic response. In these cases, the term \emph{negative index of refraction} cannot be used, but the related effects, negative refraction and backward propagating waves, are present \cite{Dumelow1993,Belov2003}. Such situation happens for the $p$-polarized reflection from an anisotropic material. The expression for $\kappa_p$ in Eq. \eqref{kappa p s} reveals that anomalous refraction can occur if the real parts of the dielectric function values, $\varepsilon_x$ and $\varepsilon_z$, have opposite sign. This occurs naturally in an anisotropic material in the vicinity of $\omega_\text{LO}$. As elaborated by Belov \cite{Belov2003}, $\varepsilon_x >0$, $\varepsilon_z < 0$ will lead to negative refraction in terms of Poynting vector $\boldsymbol{S}$, but positive (forward-wave) solution for $\kappa_p$. Vice versa, negative $\kappa_p$, characterized by backward wave propagation, will lead to positive refraction in terms of energy flow, when the condition is reversed, $\varepsilon_x<0$, $\varepsilon_z>0$.

Indeed, such a negative refraction effect has been observed for several crystals, e.g. Quartz \cite{Silva2010} or MgF$_\text{2}$ \cite{Macedo2014} and it appears also in the anisotropic effective response of dielectric heterostructures \cite{Hoffman2007} and their ellipsometric spectra \cite{Humlicek2011}. Furthermore, such phenomena attracted attention for possible photonics applications with engineered hyperbolic meta-materials \cite{Poddubny2013}.

Although the negative refraction is not observed in our experiments directly, the ellipsometric spectra reveal some corresponding features that we will discuss for the case of orthorhombic YAlO$_\text{3}$, having access to all permutations of the sequence of the anisotropic $\omega_\text{LO}$ frequencies.

\section{Results and discussion}

Figs. \ref{YAO_a_axis_response}a, \ref{YAO_b_axis_response}a and \ref{YAO_c_axis_response}a show the infrared ellipsometry spectra at room temperature measured on the three different high symmetry surfaces for 6 high symmetry configurations. Fig. \ref{YAO_a_axis_response}a shows the spectra for the $b$-cut and $c$-cut surfaces with the $a$-axis parallel to the $x$-axis (plane of incidence) of the ellipsometer ($abc$, $acb \leftrightarrow xyz$). Fig. \ref{YAO_b_axis_response}a displays the corresponding spectra with the $b$-axis parallel to $x$ for the $a$-cut and $c$-cut surfaces ($bac$, $bca \leftrightarrow xyz$) and Fig. \ref{YAO_c_axis_response}a with the $c$-axis axis parallel to $x$ for $a$-cut and $b$-cut surfaces ($cab$, $cba \leftrightarrow xyz$).

\subsection{Pseudo-isotropic response}
\begin{figure*}
\includegraphics[width=0.95\textwidth]{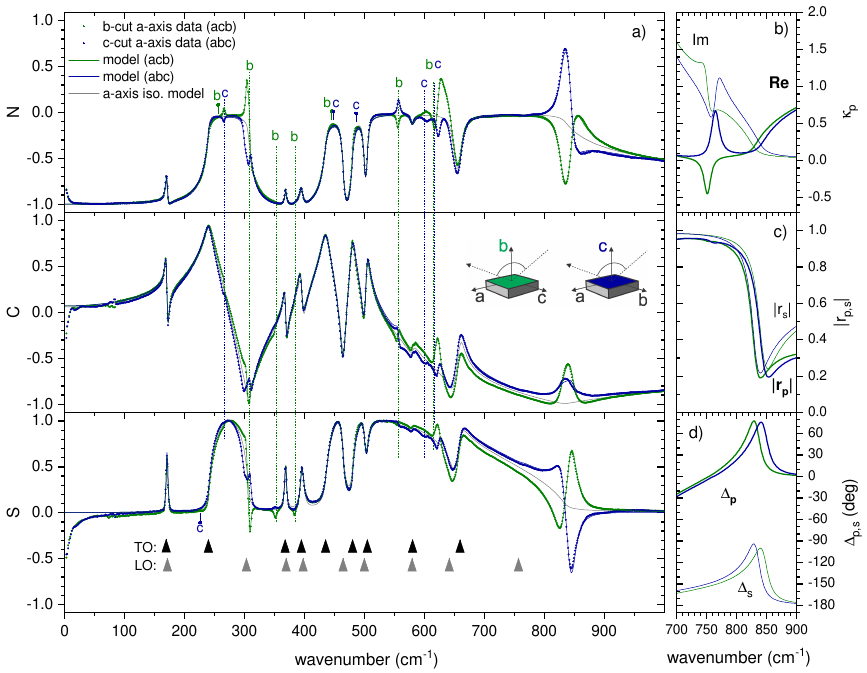}
\caption{\label{YAO_a_axis_response} a) Far-infrared ellipsometry spectra of the Mueller matrix elements $N$, $C$, $S$ of YAlO$_\text{3}$ at \SI{300}{\kelvin} with angle of incidence $\varphi = \SI{75}{\degree}$ measured on $b$-cut (green points) and $c$-cut (blue points) surfaces with the $a$-axis parallel to the plane of incidence. Sketches of the measurement configurations are shown in the inset of the middle panel. Also shown are the best fits with the anisotropic model (blue and green lines) and, for comparison and as guides to the eye, the spectra for a fictive sample with an isotropic response that have been calculated with the parameters of $\varepsilon_a$ (gray lines). Marked by arrows on the bottom panel are the positions of $\omega_\text{TO}$ (black arrows) and $\omega_\text{LO}$ (gray arrows) of $\varepsilon_a$. Dashed vertical lines highlight some of the pronounced anisotropy features in the vicinity of $\omega_\text{LO}$ in the $b$-axis response (green dashed line) and the $c$-axis response (blue dashed line). Solid pin symbols show positions of strong $\omega_\text{TO}$ in the $b$-axis response (green pin) and the $c$-axis response (blue pin). b) Calculated geometric $z$-component of the refracted wavevector for p-polarized light, Re\{$\kappa_p$\} (thick lines) and Im\{$\kappa_p$\} (thin lines), with the colors matching the configuration of the experiment. c) Calculated absolute values of the Fresnel coefficients $|r_p|$ (thick lines) and $|r_s|$ (thin lines). d) Calculated argument of the complex Fresnel coefficients $\arg(r_p) = \Delta_p$ (thick lines) and $\arg(r_s) = \Delta_s$ (thin lines).}
\end{figure*}

\begin{figure*}
\includegraphics[width=0.95\textwidth]{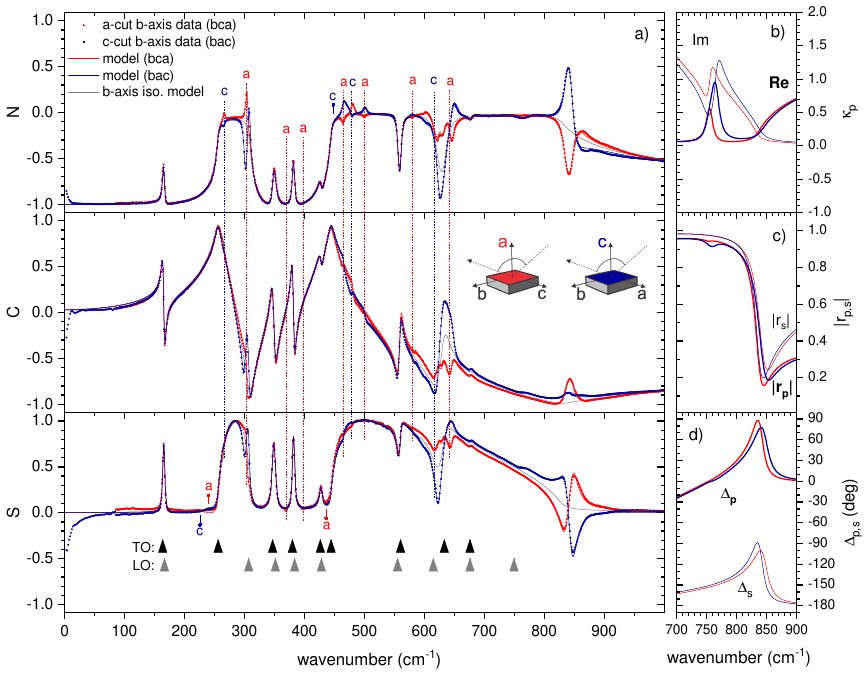}
\caption{\label{YAO_b_axis_response} a) Far-infrared response along the $b$-axis of YAlO$_\text{3}$ at \SI{300}{\kelvin} with $\varphi = \SI{75}{\degree}$ measured on $a$-cut (red points) and $c$-cut (blue points) surfaces (see the sketches in the inset of the middle panel) and expressed in terms of the Mueller matrix elements $N$, $C$, $S$. Also shown are the best fits with the anisotropic model (red and blue lines) as well as the simulated response of a fictive isotropic sample with the $\varepsilon_b$ parameters as obtained from the anisotropic model (gray lines), as a guide to the eye. Arrows on the bottom panel mark the positions of $\omega_\text{TO}$ (black arrows) and $\omega_\text{LO}$ (gray arrows) of $\varepsilon_b$. Dashed vertical lines highlight some of the pronounced anisotropy features in the vicinity of $\omega_\text{LO}$ in the $a$-axis response (red dashed line) and the $c$-axis response (blue dashed line). Solid pin symbols show positions of strong $\omega_\text{TO}$ in the $a$-axis response (red pin) and the $c$-axis response (blue pin). b) Calculated geometric $z$-component of the refracted wavevector for $p$-polarized light, Re\{$\kappa_p$\} (thick lines) and Im\{$\kappa_p$\} (thin lines), with colors matching the configuration of the experiment. c) Calculated absolute values of the Fresnel coefficients $|r_p|$ (thick lines) and $|r_s|$ (thin lines). d) Calculated argument of the complex Fresnel coefficients $\arg(r_p) = \Delta_p$ (thick lines) and $\arg(r_s) = \Delta_s$ (thin lines).}
\end{figure*}

\begin{figure*}
\includegraphics[width=0.95\textwidth]{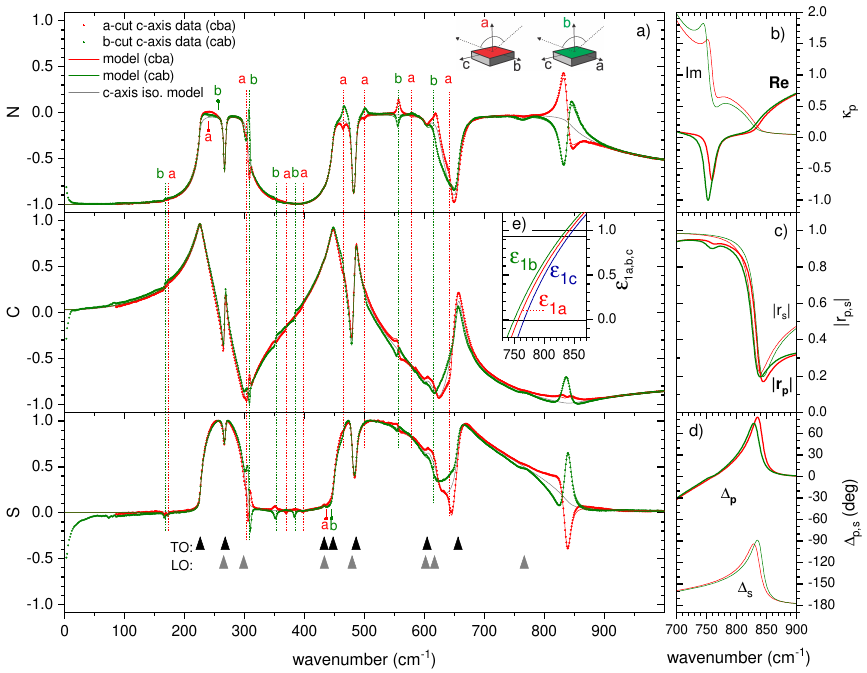}
\caption{\label{YAO_c_axis_response} a) Far-infrared response along the c-axis of YAlO$_\text{3}$ at \SI{300}{\kelvin} with $\varphi = \SI{75}{\degree}$ as measured on $a$-cut (red points) and $b$-cut (green points) surfaces and shown by the Mueller matrix elements $N$, $C$, $S$. Also displayed are the best fits with the anisotropic model (red and green lines) and the response of a fictive isotropic sample simulated with the $\varepsilon_c$ parameters (gray lines), as a guide to the eye. The inset (e) of the middle panel shows a comparison of the model of the real part of dielectric functions $\varepsilon_a$, $\varepsilon_b$, and $\varepsilon_c$. Arrows on the bottom panel mark the positions of $\omega_\text{TO}$ (black arrows) and $\omega_\text{LO}$ (gray arrows) of $\varepsilon_c$. Dashed vertical lines highlight some of the pronounced anisotropy features in the vicinity of $\omega_\text{LO}$ in the $a$-axis response (red dashed line) and the $b$-axis response (green dashed line). Solid pin symbols show positions of strong $\omega_\text{TO}$ in the $a$-axis response (red pin) and the $b$-axis response (green pin).  b) Calculated geometric $z$-component of the refracted wavevector for $p$-polarized light, Re\{$\kappa_p$\} (thick lines) and Im\{$\kappa_p$\} (thin lines), with colors matching the configuration of the experiment. c) Calculated absolute values of the Fresnel coefficients $|r_p|$ (thick lines) and $|r_s|$ (thin lines). d) Calculated argument of the complex Fresnel coefficients $\arg(r_p) = \Delta_p$ (thick lines) and $\arg(r_s) = \Delta_s$ (thin lines).}
\end{figure*}

In addition to the measured spectra (colored symbols), Figs. \ref{YAO_a_axis_response}a, \ref{YAO_b_axis_response}a and \ref{YAO_c_axis_response}a also show the best fits with the anisotropic model (colored lines) and - for comparison and as guides to the eye - the simulated ellipsometric spectra (using the fit parameters derived with the anisotropic model) for the fictive scenario of an isotropic response for $\varepsilon_a$, $\varepsilon_b$, and $\varepsilon_c$, respectively (gray lines). These isotropic model lines serve to identify the spectral ranges where the anisotropy has a clear effect on the experimental data. They make it apparent that in large parts the ellipsometric spectra are dominated by the dielectric tensor component along the axis $x$ direction. This circumstance is also evident from Eq. \eqref{r p s}, where $r_p$ depends explicitly on $\varepsilon_x$, due to the matching condition of the in-plane components of the electromagnetic field on the interface. Moreover, in Eq. \eqref{kappa p s} the value of $\varepsilon_z$ appears only as the denominator of $\xi^2/\varepsilon_z$ such that $\kappa_p \approx \sqrt{\varepsilon_x}$ if $\varepsilon_z \gg 1$.

Nevertheless, the measured far infrared ellipsometric spectra depend - to some extent - on all three components of the orthorhombic dielectric tensor. In agreement, the data and the anisotropic model reveal some clear deviations from the virtual isotropic model that occur in the vicinity of the LO frequencies of the response functions in the $y$ and $z$ directions (marked by vertical dashed lines). Additional minor deviations can be attributed to the strongest TO modes of the $y$ response (marked by solid pin symbols). On the figures \ref{YAO_a_axis_response}a, \ref{YAO_b_axis_response}a and \ref{YAO_c_axis_response}a we marked only those mode frequencies where noticeable anisotropic features are observed. These are typically the stronger and sharper LO$_y$ and LO$_z$ modes, and only the strongest TO$_y$ modes. For the reason discussed at the end of previous paragraph, strong TO$_z$ modes have no effect on the spectra. In the Supplemental Material \cite{SupMat} we present alternative plots, with spectra paired by the sample surface cut, i.e. spectra obtained by rotation of a single sample, and with markers for all 25 TO and 25 LO modes.

Here, at first, we discuss the overall pseudo-isotropic behavior of the NCS spectra, while the anisotropy related features will be discussed further below. The frequencies of the TO$_x$ and LO$_x$ phonon modes in the spectra of $\varepsilon_a$, $\varepsilon_b$, and $\varepsilon_c$ (as fitted with the anisotropic model) are marked by black and gray arrows in the bottom panels of Figs. \ref{YAO_a_axis_response}a, \ref{YAO_b_axis_response}a and \ref{YAO_c_axis_response}a, respectively. Note that our definition of $N=-\cos 2 \Psi$ differs in sign with respect to the convention used in most textbooks ($N=\cos 2 \Psi$). This sign change has the advantage that i) the value of $N$ equals the Mueller matrix element $N=m_{21}=m_{12}$, ii) the shape of the $N$-spectra resembles that of ellipsometric angle $\Psi$ (which is widely used for presenting ellipsometric data), or $\tan \Psi$, and iii) their shape is also reminiscent of the normal incidence reflectivity, $R$. Accordingly, the so-called Reststrahlen bands are clearly visible in the spectra of $N$ as maxima that span the region of the TO-LO splitting of the phonon modes, at which $N \rightarrow 0$ ($\Psi \rightarrow \SI{45}{\degree}$ and $R \rightarrow 1$). More precisely, the apparent Reststrahlen band ends at a frequency $\omega_{\text{LO}^*}$, which depends on the angle of incidence and will be properly introduced in the next section, as it shows strongly anisotropic behavior.

The Mueller matrix element $S$ is complementary to $N$, since its value approaches zero ($S \rightarrow 0$) between the Reststrahlen bands, that is, for range between LO$_x$ and the following TO$_x$ frequency. The Mueller matrix element $C$ shows a characteristic zigzag shape with maxima at the TO$_x$ and minima at the LO$_x$ frequencies.

From there it can be inferred that the ellipsometric data in NCS form are equally well represented by the dielectric function $\varepsilon$ as they are by its inverse $1/\varepsilon$. The factorized form of Eq.\eqref{FPSQ diel func} allows for straightforward inversion. We have also simulated the spectra using additive, harmonic oscillator approximation (HOA) model. Since YAlO$_\text{3}$ does not show strongly asymmetric phonon line-shapes, the additive HOA model for $\varepsilon$ represents the data qualitatively well. For the same reason, the third tested alternative, an additive HOA model for $1/\varepsilon$, produces comparably good fit. Nevertheless, the utilized four-parameter model shows the best performance, particularly on the strong TO modes at \SI{240}{\per\centi\meter}, \SI{256}{\per\centi\meter}, and \SI{268}{\per\centi\meter} in the $a$-, $b$-, and $c$-axis response, respectively, and around the highest LO edge as will be discussed further below.  

Towards very low frequencies in the THz regime, the measured spectra start to exhibit significant deviations from those of the model curves that are caused by diffraction effects \cite{Humlicek2004}. In particular, the spectra of $S$ exhibit a steep downturn below approximately \SI{100}{\per\centi\meter}. The spectra of $N$ and $C$ are less strongly affected by the diffraction and show clear deviations only below \SI{20}{\per\centi\meter}. Besides these diffraction artifacts (which prevail below \SI{100}{\per\centi\meter}), the $a$-cut $b$-axis data of the element $S$, in Fig. \ref{YAO_b_axis_response}a, are slightly deviating from the expected model values below \SI{250}{\per\centi\meter}. Likely, this experimental artifact is caused by a backside reflection or light that is otherwise scattered (but still reaches the detector). Otherwise, the model shows excellent agreement with the measured data and thus allows us to identify and interpret the observed features.

\subsection{Anisotropy features}
With the help of the above-described model, we can now discuss the signatures of the anisotropy in the ellipsometry spectra of Fig. \ref{YAO_a_axis_response} to \ref{YAO_c_axis_response}. The anisotropic features generally occur in the vicinity of the LO modes of those crystal axes that are oriented along the $y$- and $z$-direction of the coordinate system of the ellipsometer (see Fig. \ref{YAO_schema}b). These are marked by the vertical dashed lines across the NCS panels in Figs. \ref{YAO_a_axis_response}a, \ref{YAO_b_axis_response}a, and \ref{YAO_c_axis_response}a. Strong TO$_y$ modes (oriented along the $y$-direction) lead to visible changes in the spectra only when in the vicinity of a strong TO$_x$ mode. We have marked position of the strongest TO$_y$ in the plots by solid pin symbols.

The TO$_y$ related deviations are most apparent on the leading TO edge of the $N$ and $S$ spectra due to strong TO phonons at \SI{240}{\per\centi\meter}, \SI{256}{\per\centi\meter}, and \SI{268}{\per\centi\meter} in the $a$-, $b$-, and $c$-axis response, respectively. Whenever TO$_y$ mode follows TO$_x$, the flat part of the $N\approx0$ band shows elevation (shoulder) with $N>0$ spanning between TO$_x$ and TO$_y$. Such shoulder is visible in the $abc$ spectrum on figure \ref{YAO_a_axis_response}a, and - more clearly - in the $cba$ and $cab$ spectra on figure \ref{YAO_c_axis_response}a. On the other hand, when strong TO$_y$ mode appears below the frequency of TO$_x$, the flat part of the $S\approx0$ band on the foot of the TO edge shows a dip, $S<0$, spanning TO$_y$ to TO$_x$, which is visible on figure \ref{YAO_a_axis_response}a for $acb$ and on figure \ref{YAO_b_axis_response}a for $bca$ and $bac$ configurations. Corresponding features occur also near the TO edges in the 430 to \SI{490}{\per\centi\meter} range, but there they are not quite visible on the plots. These phenomena become much more clear in the low temperature spectra, where phonon line-shapes are sharper. In the Supplemental Material \cite{SupMat} we have plotted the 10 K model lines for added clarity.

Up to approximately \SI{600}{\per\centi\meter} the spectra follow mostly the guide-to-the-eye isotropic model, the anisotropy only gives raise to some rather localized features near the TO$_y$, LO$_y$ and LO$_z$  frequencies. This changes above \SI{600}{\per\centi\meter} where the spectra reveal strong anisotropy features, in particular, in the ranges from \SIrange{600}{650}{\per\centi\meter} and from \SIrange{800}{900}{\per\centi\meter}. The response in the range from \SIrange{600}{650}{\per\centi\meter} is somewhat complicated by the overlap of the TO and LO modes from all three directions, with relatively strong LO modes but rather broad and thus weak TO modes that are superimposed on a weak dielectric background. Notably, the strong anisotropy features around \SI{850}{\per\centi\meter} appear well above the highest LO frequencies that are similar for all three axes and occur at 750-\SI{770}{\per\centi\meter} (see Fig. \ref{YAO_c_axis_response}e, inset in the middle panel of Fig. \ref{YAO_c_axis_response}a). Such features, above the highest LO frequency, are commonly observed in the ellipsometric spectra of orthorhombic and tetragonal systems \cite{Schubert2000,Schoeche2013}. They occur at the end of the Reststrahlen band, which in normal-incidence reflectivity experiment coincides with $\omega_\text{LO}$ (where $\varepsilon \rightarrow 0$), but for an oblique angle of incidence, $\varphi$, is shifted to the frequency at which $\varepsilon \sim \sin^2\varphi$, i. e. to the point where the $z$-component of the wavevector vanishes (see Eq. \eqref{kappa p s}). In the literature this frequency is commonly denoted as $\omega_{\text{LO}^*}$. The shift of the Reststrahlen band edge with the angle of incidence occurs also for isotropic materials, but without the anisotropy effects.

The strong features that occur around the $\omega_{\text{LO}^*}$ point in the response of anisotropic samples arise from the difference between the $\omega_{\text{LO}^*}$ frequencies of the dielectric tensor components that are aligned with the $x$, $y$, and $z$ directions. Specifically, the strong peaks and dips appear around the frequencies at which the dielectric functions $\varepsilon_a$, $\varepsilon_b$ or $\varepsilon_c$ are crossing the value of $\sin^2\varphi$ ($\equiv \xi^2$ in Eq. \eqref{kappa p s}). For the presented data, which have been obtained for an angle of incidence $\varphi = \SI{75}{\degree}$, this corresponds to the condition  $\sin^2\varphi = 0.933$ that has been marked with a horizontal line in Fig. \ref{YAO_c_axis_response}e and yields estimates of $\omega_{a,\text{LO}^*} \approx \SI{834}{\per\centi\meter}$, $\omega_{b,\text{LO}^*} \approx \SI{829}{\per\centi\meter}$, and $\omega_{c,\text{LO}^*} \approx \SI{841}{\per\centi\meter}$. Also marked are the frequencies at which the dielectric functions reach unity and thus equal that of the ambient. This matching with the dielectric constant of the ambient occurs at  \SI{842}{\per\centi\meter}, \SI{836}{\per\centi\meter} and \SI{848}{\per\centi\meter} for $\varepsilon_a$, $\varepsilon_b$ and $\varepsilon_c$, respectively. At these frequencies, the Fresnel coefficients acquire their extremal values, as well as their ratio $\rho = r_p/r_s$, and subsequently $\Psi$, $\Delta$, or $N$, $C$, $S$. The strong spectral features that originate from the contrast between $r_p$ and $r_s$, are governed here by the divergence of the $\xi^2/\varepsilon_z$ term in Eq. \eqref{kappa p s} which strongly impacts $r_p$ (but not $r_s$.). This is unlike the pseudo-isotropic parts of the spectra below \SI{600}{\per\centi\meter} for which $r_p$ is dominated by $\varepsilon_x$. Note that in both cases the $y$-response solely affects $r_s$.

As a result, in the vicinity of $\omega_{\text{LO}^*}$ it is the contrast between $\varepsilon_z$ and $\varepsilon_y$ that gives rise to the anisotropy features in the ellipsometry spectra that appear in the same frequency range but show opposite trends (minima versus maxima) for the pairs of different configurations ($abc$ versus $acb$ in Fig. \ref{YAO_a_axis_response}a, $bac$ versus $bca$ in Fig. \ref{YAO_b_axis_response}a and $cab$ versus $cba$ in Fig. \ref{YAO_c_axis_response}a. Note that this mutually opposite behavior is a general characteristic for all the LO anisotropy features of the infrared spectra.

This effect can be understood as follows: Fig. \ref{YAO_a_axis_response}b displays the $z$-component of the geometrical wavevector of the $p$-polarized light, $\kappa_p$, for the two configurations $abc$ and $acb$, calculated using the values of the best fit with the anisotropic model. For the moment we skip the feature occurring at the actual $\omega_\text{LO}$, which will be discussed later. In general, the position of $\omega_{\text{LO}^*}$ coincides with the minimum of the absolute value of $\kappa_p$. For the $b$-cut surface ($b \parallel z$) the minimum of $|\kappa_p|$ thus occurs at $\omega_{b,\text{LO}^*}$ and for the $c$-cut sample ($c \parallel z$) at $\omega_{c,\text{LO}^*}$. The corresponding $s$-component, $\kappa_s$ (not plotted), has an overall similar shape with a minimum at the $\omega_{\text{LO}^*}$ of the $y$-axis response, but does not contain the features related to $\omega_\text{LO}$. The magnitudes of the corresponding complex Fresnel coefficients $|r_p|$ and $|r_s|$ are plotted in Fig. \ref{YAO_a_axis_response}c, and their phases $\Delta_p$ and $\Delta_s$ on Fig. \ref{YAO_a_axis_response}d.

The minima of $|\kappa_{p,s}|$ at $\omega_{\text{LO}^*}$ also give rise to characteristic features in spectra phase shifts, $\Delta_p$ and $\Delta_s$. For the $b$-cut surface, $\Delta_p$ and $\Delta_s$ have fairly sharp maximum at $\omega_{b,\text{LO}^*} = \SI{829}{\per\centi\meter}$ and $\omega_{c,\text{LO}^*} = \SI{841}{\per\centi\meter}$, respectively. For the $c$-cut surface, this order is inverted, i.e. the maxima of $\Delta_p$ and $\Delta_s$ occur at $\omega_{c,\text{LO}^*} = \SI{841}{\per\centi\meter}$ and $\omega_{b,\text{LO}^*} = \SI{829}{\per\centi\meter}$, respectively.  This splitting of the maxima of the phase shifts $\Delta_p$ and $\Delta_s$, for which the order is inverted between $b$- and $c$-cut surfaces, gives rise to pronounced features (dips and kinks) in the ellipsometric angle Delta ($\Delta = \Delta_p - \Delta_s$), and consequently in the Mueller matrix element $S$ in the lower panel of Fig. \ref{YAO_a_axis_response}a. For the latter, the extrema are slightly shifted, i.e. they occur slightly below $\omega_{b,\text{LO}^*} = \SI{829}{\per\centi\meter}$ and slightly above $\omega_{c,\text{LO}^*} = \SI{841}{\per\centi\meter}$.

For the interpretation of the corresponding feature in $N$, we recall that $N$ is closely related to the ellipsometric angle $\Psi$ ($N=-\cos2\Psi$), and $\tan\Psi = |r_p|/|r_s|$. The extremum in $N$ coincides here with the minima of $|r_p|$ and $|r_s|$ (Fig. \ref{YAO_a_axis_response}c) which occur at the point at which the contrast between dielectric constant of the sample and the ambient vanishes. For $|r_s|$ this occurs as the $y$-axis response crosses unity, i.e. for the $c$-cut surface ($abc$) at \SI{836}{\per\centi\meter} for $\varepsilon_b$ and for the $b$-cut surface (\textit{acb}) at \SI{848}{\per\centi\meter} for $\varepsilon_c$. Note that a finite absorption broadens these $|r_s|$ minima and shifts them to slightly higher frequencies.

The coefficient $|r_p|$ depends in general on both the $x$-axis and $z$-axis responses (see Eq. \ref{kappa p s}, \ref{r p s}). However, since in the relevant range both of them are close to unity, the minimum of $|r_p|$ is more sensitive to the $z$-axis response and appears slightly above the point where the $z$-axis response crosses unity. Note that this $|r_p|$ minimum also represents one of the solutions for the general Brewster condition \cite{Dumelow1993}.

For the sake of completeness, we have also plotted the corresponding $\kappa_p$, $|r_p|$ and $|r_s|$ spectra for the other sample orientations and measurement configurations in Fig. \ref{YAO_b_axis_response}b-d and \ref{YAO_c_axis_response}b-d. The resulting anisotropy effects in the NCS spectra are analogous to the ones that we have discussed above for the $abc$ and $acb$ configurations that are displayed in Fig. \ref{YAO_a_axis_response}a. They simply involve different permutations of the $a$-, $b$- or $c$-components of the relevant critical frequencies.

The above discussion about the large anisotropy features in the ellipsometric response in the range from \SIrange{800}{900}{\per\centi\meter} can be summarized as follows: The features in the Mueller matrix element $N$ (or in the ellipsometric angle $\Psi$) are related to the anisotropy of the minima in $|r_p|$ and $|r_s|$ that occur if the dielectric function of the sample matches that of the ambient (i.e. $\varepsilon \rightarrow 1$ in case of vacuum ambient). The corresponding features in the Mueller matrix element $S$ (or in $\Delta$) depend on the actual anisotropy of $\omega_{\text{LO}^*}$ ($|\kappa| \rightarrow 0$, onset of total reflection), which at $\varphi = \SI{75}{\degree}$ yields the condition $\varepsilon = 0.933$. Note that both kind of features get broadened and slightly shifted by the nonzero imaginary parts of the response functions. Moreover, since these critical points tend to be rather close in frequency, the different contributions to the anisotropy effects in NCS are partially merged, especially if the angle of incidence is large with $\xi^2 \rightarrow 1$. Measurements at a lower angle of incidence would allow one to better separate the features appearing at $\omega_{\text{LO}^*}$ and at $\varepsilon \rightarrow 1$. In both cases, however, these anisotropy features are directly related to the differences of the z-axis and y-axis response functions ($\varepsilon_z$ and $\varepsilon_y$). Notably, near $\omega_{\text{LO}^*}$ these features are almost independent of $\varepsilon_x$ – although at lower frequencies, the $x$-axis response dominates the ellipsometry spectra.

Note that the smaller anisotropy effects that appear near the $\omega_\text{LO}$ frequencies in the range below \SI{600}{\per\centi\meter} (marked by the vertical dashed lines in Figs. \ref{YAO_a_axis_response}a, \ref{YAO_b_axis_response}a, \ref{YAO_c_axis_response}a) have the same origin as described above for the \SIrange{800}{900}{\per\centi\meter} range. The main difference concerns the separation between $\omega_\text{LO}$, $\omega_{\text{LO}^*}$ and the point of $\varepsilon \rightarrow 1$ that becomes much smaller.

\subsection{Negative refraction}
Another interesting feature concerns an anomaly in the geometrical $z$-component of the refracted wavevector of the p-polarized beam, $\kappa_p$, that occurs in the vicinity of the highest LO mode, i.e. near the zero crossing of the real part of $\varepsilon$ around \SI{750}{\per\centi\meter} (see Figs. \ref{YAO_a_axis_response}b, \ref{YAO_b_axis_response}b and \ref{YAO_c_axis_response}b, see also inset Fig. \ref{YAO_c_axis_response}e). In the experimental data, the anisotropy of this highest LO mode with $\omega_{b,\text{LO}}<\omega_{a,\text{LO}}<\omega_{c,\text{LO}}$ gives rise to a small drop in $N$ (or $\Psi$) and corresponding feature in $C$ and $S$.

Notably, this feature appears well below $\omega_{\text{LO}^*}$ and thus within the Reststrahlen band, for which in case of an isotropic sample only an evanescent wave (but not a refracted wave) could exist. Such a refracted wave is however possible for the anisotropic case of YAlO$_\text{3}$ for which the expression for $\kappa_p$ in Eq. \eqref{kappa p s} has a non-evanescent solution in the range where $\varepsilon_x$ and $\varepsilon_z$ have opposite signs. For the condition $\varepsilon_x < 0$, $\varepsilon_z >0$, there exists a strongly attenuated, but propagating wave that has an unconventional refraction. The real part of the $z$-component of the wave vector $\kappa_p$, which determines the direction of the phase propagation, has a negative value such that it describes a backward traveling wave (the $x$-component, $\xi$, of the wave vector remains positive). Nevertheless, for the energy flow the refraction remains positive since the $x$- and the $z$-components of the Poynting vector are positive.

For the opposite case of  $\varepsilon_x > 0$ and $\varepsilon_z < 0$, there is negative refraction effect: the $x$-component of the Poynting vector becomes negative, although the real part of $\kappa_p$ is positive and the waves are traveling forward from the interface to the bulk \cite{Belov2003}. The Figs. \ref{YAO_a_axis_response}b, \ref{YAO_b_axis_response}b, \ref{YAO_c_axis_response}b show the $\kappa_p$ spectra calculated with the parameters of the best-fit model. In Fig. \ref{YAO_a_axis_response}b, the solution for the $abc$-configuration has a positive Re\{$\kappa_p$\} around \SI{770}{\per\centi\meter} – leading to a forward wave but a negative energy refraction. The $acb$-configuration shows a negative Re\{$\kappa_p$\} around \SI{750}{\per\centi\meter} – corresponding to a backward wave with positive energy flow. In Fig. \ref{YAO_b_axis_response}b, the $bac$- and $bca$-configurations have positive Re\{$\kappa_p$\} and negative refraction. Finally, in Fig. \ref{YAO_c_axis_response}b, both the $cab$- and $cba$-configurations have negative Re\{$\kappa_p$\} and positive refraction in the discussed spectral range.

The above discussion raises the question whether based on the ellipsometric data one can tell the difference between the cases of negative and positive refraction. The answer is that the ellipsometry data do not allow on to distinguish between these two cases. Indeed, it turns out that the change in the sign of the real part of $\kappa_p$ does not give rise to any characteristic feature in $r_p$ and the resulting Mueller matrix elements $N$, $C$, $S$ that could serve as indicator. This is evident from the comparison of the real part of the $\kappa_p$ curve for the \textit{bac} configuration in Fig. \ref{YAO_b_axis_response}b (with positive values) with that of the \textit{cab}-configuration in Fig. \ref{YAO_c_axis_response}b (with negative values) and an inspection of the resulting NCS spectra in Figs \ref{YAO_b_axis_response}a and \ref{YAO_c_axis_response}b which exhibit no characteristic difference in the vicinity of the high $\omega_\text{LO}$.

\begin{figure}
\includegraphics[width=0.475\textwidth]{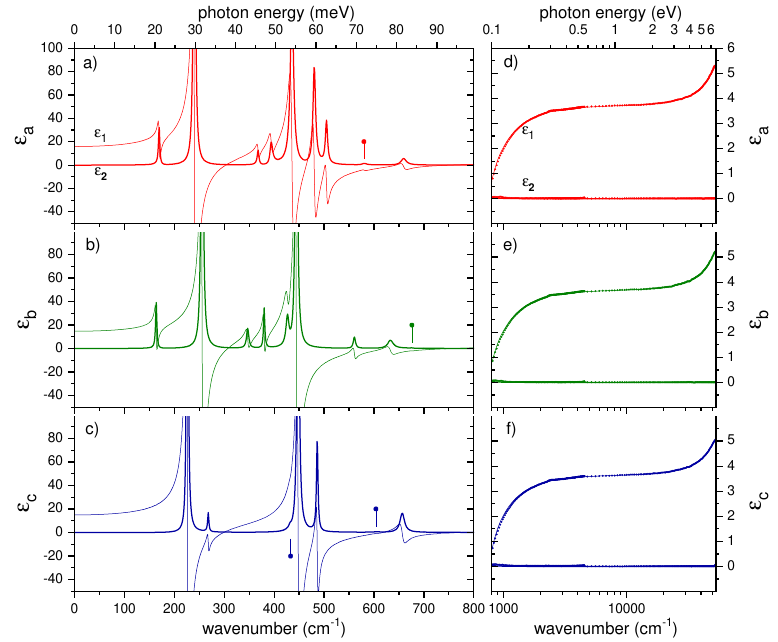}
\caption{\label{YAO diel tensor} Diagonal elements of the dielectric tensor along the three main crystal axes of orthorhombic YAlO$_\text{3}$. The spectra represent the best fit with an anisotropic model to the ellipsometry data taken at room temperature for six different configurations with respect to the surface cut and the mutual orientation of the crystal axes. Panels a), b), and c) show the far-infrared spectra of $\varepsilon_a$, $\varepsilon_b$ and $\varepsilon_c$, respectively. Thin lines represent the real part of the dielectric function, $\varepsilon_1$, thicker lines the imaginary part, $\varepsilon_2$. The positions of the weak phonons are marked with the pin symbols. Panels d), e), and f) show the extension of the model in the MIR/NIR/VIS/UV ranges on a logarithmic energy scale. Lines represent an anisotropic Cauchy model. Symbols show the result of an anisotropic point-by-point fit.}
\end{figure}

\subsection{Dielectric response along the main crystal axes of YAlO$_\text{3}$}
Figs. \ref{YAO diel tensor}a, b and c display the far-infrared dielectric response along the main crystal axes of orthorhombic YAlO$_\text{3}$ that represent the diagonal elements, $\varepsilon_a$, $\varepsilon_b$ and $\varepsilon_c$, of the dielectric tensor. They have been obtained from a simultaneously fitting with an anisotropic model to the room temperature ellipsometric data for six different configurations of the surface cuts and relative orientations of the crystal axes. The corresponding data and model curves have been presented in Figs. \ref{YAO_a_axis_response}a, \ref{YAO_b_axis_response}a, and \ref{YAO_c_axis_response}a. The model contains factors that describe the response of the infrared active TO and LO phonons, as given in Eq.\eqref{FPSQ diel func}. With this model, we identified 9 phonons ($B_{1u}$) in the $a$-axis response, $\varepsilon_a$, 9 phonons ($B_{3u}$) in the $b$-axis response, $\varepsilon_b$, and 7 phonons ($B_{2u}$) in the $c$-axis response, $\varepsilon_c$, as predicted from theory \cite{Vali2007,Dewo2020}. Several of the TO phonons shown in Fig. \ref{YAO diel tensor} are rather weak, their position has therefore been highlighted with the pin symbols.

\begin{table*}
\caption{\label{IR YAO params RT table} Infrared model parameters of the anisotropic dielectric response of YAlO$_\text{3}$ at room temperature. The fit parameters are:
$\omega_\text{TO}$, $\gamma_\text{TO}$, $\omega_\text{LO}$,  $\gamma_\text{LO}$, position and broadening of the transversal and longitudinal phonon modes, respectively, along the three orthogonal axes of the crystal in $Pbnm$ notation: $a$ ($B_{1u}$ symmetry), $b$ ($B_{3u}$ symmetry) and $c$ ($B_{2u}$ symmetry). In each axis we also fit the asymptotic value of dielectric constant above the infrared phonon range, $\varepsilon_{\infty}$, see eq. \eqref{FPSQ diel func}. Values $\Delta \varepsilon$ are the contributions of each TO phonon mode to the static dielectric constant, $\varepsilon(0)$, calculated from the TO - LO splitting of each mode. The displayed fit parameters are rounded with respect to the first significant digit of the corresponding estimated error. Parameter values marked with an asterisk were not fitted.}
\begin{ruledtabular}
\begin{tabular}{ccc}
\resizebox{0.32\textwidth}{!}{
\begin{tabular}{ccccc}
\multicolumn{5}{c}{$a$-axis ($B_{1u}$)}\\
$\omega_{\mathrm{TO}} (\mathrm{cm^{-1}})$ & $\gamma_{\mathrm{TO}} (\mathrm{cm^{-1}})$ & $\omega_{\mathrm{LO}} (\mathrm{cm^{-1}})$ & $\gamma_{\mathrm{LO}} (\mathrm{cm^{-1}})$ & $\Delta\varepsilon$ \\
\hline
169.8	& 3.2	& 172.0 	& 2.8  & 0.61\\
240.3	& 4.0	& 303.4 	& 3.5  & 7.11\\ 
368.0	& 4.1	& 369.9 	& 3.7  & 0.14\\
393.9	& 6.2	& 397.0 	& 7.6  & 0.29\\
435.8	& 7.5	& 464.4 	& 5.3  & 2.57\\
480.3	& 6.1	& 500.1 	& 5.8  & 1.07\\
504.9	& 5.2	& 579   	& 9    & 0.37\\ 
580  	& 10 	& 642.6 	& 9.9  & 0.02\\
659.2	& 12.4	& 757.3 	& 13.9 & 0.11\\
\hline
$\varepsilon_{\infty}$ & & & & 3.61 \\
$\varepsilon(0)$ & & & & 15.90  
\end{tabular}} &
\resizebox{0.32\textwidth}{!}{
\begin{tabular}{ccccc}
\multicolumn{5}{c}{$b$-axis ($B_{3u}$)}\\
$\omega_{\mathrm{TO}} (\mathrm{cm^{-1}})$ & $\gamma_{\mathrm{TO}} (\mathrm{cm^{-1}})$ & $\omega_{\mathrm{LO}} (\mathrm{cm^{-1}})$ & $\gamma_{\mathrm{LO}} (\mathrm{cm^{-1}})$ & $\Delta\varepsilon$ \\
\hline
163.9 	& 3.0 	& 167.0	& 2.7   & 0.73\\
255.9 	& 4.5 	& 307.1	& 4.2   & 5.81\\
347.0 	& 5.4 	& 351.7	& 5.2   & 0.27\\
380.0 	& 3.0 	& 383.7	& 3.4   & 0.27\\
426.9   & 7.0 	& 428.6	& 6.4   & 0.40\\
444.5 	& 4.4 	& 555.4	& 4.9   & 3.54\\
560.7 	& 5.1 	& 615.1	& 9.8   & 0.09\\
632.7 	& 13.4	& 676.7	& 7.7   & 0.15\\
677*		& 7*		& 749.9	& 15.0  & 0.001\\
\hline
$\varepsilon_{\infty}$ & & & & 3.56 \\
$\varepsilon(0)$ & & & & 14.82  
\end{tabular}} &
\resizebox{0.32\textwidth}{!}{
\begin{tabular}{ccccc}
\multicolumn{5}{c}{$c$-axis ($B_{2u}$)}\\
$\omega_{\mathrm{TO}} (\mathrm{cm^{-1}})$ & $\gamma_{\mathrm{TO}} (\mathrm{cm^{-1}})$ & $\omega_{\mathrm{LO}} (\mathrm{cm^{-1}})$ & $\gamma_{\mathrm{LO}} (\mathrm{cm^{-1}})$ & $\Delta\varepsilon$ \\
\hline
226.2	& 2.6	& 265.7	& 3.3  & 7.23\\
268.1	& 2.9	& 298.8	& 4.0  & 0.17\\
432.5*	& 6*	& 432.7	& 5.9  & 0.05\\
448.0	& 5.0	& 479.3	& 4.3  & 3.09\\
486.4	& 3.5	& 603.6	& 12   & 0.57\\
606.2	& 11	& 617.0	& 8.8  &0.007\\
656.6	& 11.0	& 766.9	& 15.4 & 0.27\\
&&&\\
&&&\\
\hline
$\varepsilon_{\infty}$ & & & & 3.48 \\
$\varepsilon(0)$ & & & & 14.87  
\end{tabular}}
\end{tabular}
\end{ruledtabular}
\end{table*}

The positions, $\omega_\text{TO}$, $\omega_\text{LO}$, and broadenings, $\gamma_\text{TO}$, $\gamma_\text{LO}$, were fitting parameters of the model. The obtained best-fit parameters for the room temperature spectra are listed in Table \ref{IR YAO params RT table}. The additional fit parameter $\varepsilon_{\infty}$ represents the high-frequency dielectric constant. The oscillator strengths, $\Delta \varepsilon$, were derived Eq.\eqref{FPSQ diel func} by setting the broadening parameters to zero and comparing with equivalent additive harmonic oscillator model, see \cite{SupMat}. Finally, the $\varepsilon(0)$ represents the static dielectric constant calculated from the sum of $\varepsilon_{\infty}$ and the oscillator strengths of the phonons, $\Delta \varepsilon$.

To verify the parameterized anisotropic model discussed here, we have additionally performed a fit to the same data with an unconstrained point-by-point model of the three diagonal elements of the dielectric tensor. Such fit did not reveal any discrepancies, nevertheless,  we have plotted the resulting spectra in the Supplemental Material \cite{SupMat}.

The obtained value of the TO phonon positions listed in Table \ref{IR YAO params RT table} are in rather good agreement with the previous theoretical predictions \cite{Vali2007,Dewo2020}. Most of the TO eigenfrequencies are within \SI{3}{\per\centi\meter} of the predicted values. In some cases, the predictions differ up to \SI{10}{\per\centi\meter} from each other, while the experimental value lies in between. Only in few cases the experimental values differ by up to \SI{10}{\per\centi\meter} from theory. The phonon at \SI{432.5}{\per\centi\meter} in the c-axis response is extremely weak, in the experimental data it is only evident in the $S$-element (bottom panel of Fig. \ref{YAO_c_axis_response}a, and thus could only be identified with the help of the theory.

Overall, these findings show that YAlO$_\text{3}$ has very stable and predictable lattice dynamical properties: A single model describes very well the experimental data that have been obtained on three samples with different surface cuts. Moreover, the older and newer theoretical calculations (of TO phonon positions) agree reasonably well and match the data to high accuracy.

Figs. \ref{YAO diel tensor}d, e and f show the model dielectric functions $\varepsilon_a$, $\varepsilon_b$ and $\varepsilon_c$ for an extended energy range that spans the mid-, near-infrared, visible and ultraviolet (up to \SI{6.5}{\electronvolt}). Here, we employed two different modeling approaches. For the first type of fitting, the whole MIR to UV range has been represented by an anisotropic Cauchy dispersion (model lines in Figs. \ref{YAO diel tensor}d, e, f), while the parameters describing the phonons in the FIR range have been fixed and the FIR data have not been included in the fit. The phonon response accounts for the downturn of the real part of the dielectric function, $\varepsilon_1$, in the mid-infrared range.

These fits have been performed simultaneously on the mid-infrared data, measured in the range \SIrange{0.1}{0.5}{\electronvolt} at $\varphi = \SI{75}{\degree}$, and on the NIR-UV data in the range from \SIrange{0.5}{6.5}{\electronvolt} measured with a Woollam ellipsometer (VASE) at variable angles of incidence of $\varphi = \SI{55}{\degree}$, $\SI{65}{\degree}$, $\SI{75}{\degree}$ and $\SI{85}{\degree}$. As for the phonons in the FIR-range, the fitting has been performed on an extended data set for all the 6 high symmetry configurations.

The surface roughness, which is known to affect especially the data in the VIS/UV region, has been accounted for by introducing an isotropic 50\% Bruggeman effective medium (BEMA) layer which mixes the actual $x$-axis response of the YAlO$_\text{3}$ sample (depending on orientation) with the ambient. Such simplification is warranted here, as: i) in the ultrathin film limit, the ellipsometric experiment is only sensitive to the $x$-axis response of the film \cite{Marsik2022}, ii) the spread of the anisotropic refractive indices in the transparent range is small, and iii) BEMA is already an approximation and yields an effective thickness only. The obtained thickness values of the BEMA layers are \SI{2.8}{\nano\meter}, \SI{1.9}{\nano\meter} and \SI{4.4}{\nano\meter} for the $a$-cut, $b$-cut and $c$-cut surfaces, respectively.

\begin{table}
\caption{\label{NIR UV Cauchy param YAO table} Near-infrared to UV Cauchy model parameters of YAlO$_\text{3}$, room temperature. Calculated anisotropic dielectric constant and refractive index at \SI{2}{\electronvolt}.}
\begin{ruledtabular}
\begin{tabular}{ccc}
\resizebox{0.15\textwidth}{!}{
\begin{tabular}{ccc}
\multicolumn{3}{c}{$a$-axis}\\
$\varepsilon_{a0}$ & 3.700 &\\
$\varepsilon_{a1}$ & 0.0150 & $\mathrm{eV^{-2}}$\\
$\varepsilon_{a2}$ & 0.00057 & $\mathrm{eV^{-4}}$\\
&&\\
$\varepsilon_a$(2eV) & 3.764 &\\
$n_a$(2eV) & 1.940 &
\end{tabular}} &
\resizebox{0.15\textwidth}{!}{
\begin{tabular}{ccc}
\multicolumn{3}{c}{$b$-axis}\\
$\varepsilon_{b0}$ & 3.677 &\\
$\varepsilon_{b1}$ & 0.0133 & $\mathrm{eV^{-2}}$\\
$\varepsilon_{b2}$ & 0.00056 & $\mathrm{eV^{-4}}$\\
&&\\
$\varepsilon_b$(2eV) & 3.734 &\\
$n_b$(2eV) & 1.932 &
\end{tabular}} &
\resizebox{0.15\textwidth}{!}{
\begin{tabular}{ccc}
\multicolumn{3}{c}{$c$-axis}\\
$\varepsilon_{c0}$ & 3.625 &\\
$\varepsilon_{c1}$ & 0.0142 & $\mathrm{eV^{-2}}$\\
$\varepsilon_{c2}$ & 0.00048 & $\mathrm{eV^{-4}}$\\
&&\\
$\varepsilon_c$(2eV) & 3.684 &\\
$n_c$(2eV) & 1.920 &
\end{tabular}}
\end{tabular}
\end{ruledtabular}
\end{table}

The best-fit parameters of the Cauchy model are listed in Table \ref{NIR UV Cauchy param YAO table}. The optimized Cauchy constants $\varepsilon_{a0}$, $\varepsilon_{b0}$ and $\varepsilon_{c0}$ differ from the corresponding values of $\varepsilon_{\infty}$ listed in Table \ref{IR YAO params RT table}. While for an ideal case these parameters should be identical, we could not achieve perfect match of model and data for the entire THz to UV range.

For the second fitting approach, we have fixed the thicknesses of the effective roughness layers, (to the value derived from the Cauchy plus BEMA fit) and fitted the same extended set of ellipsometry data in a point-by-point manner with an unconstrained orthorhombic model (fitting independently the real and imaginary parts of $\varepsilon_a$, $\varepsilon_b$ and $\varepsilon_c$). The obtained dielectric functions are shown by symbols in Fig. \ref{YAO diel tensor}d, e and f. It is evident that the result of the point by-point fit agrees well with the one derived from the fitting with the parameterized model functions. Notably, no sign of energy gap edge is detected up to \SI{6.5}{\electronvolt}.

\subsection{Temperature dependence of the phonons YAlO$_\text{3}$}
\begin{figure}
\includegraphics[width=0.475\textwidth]{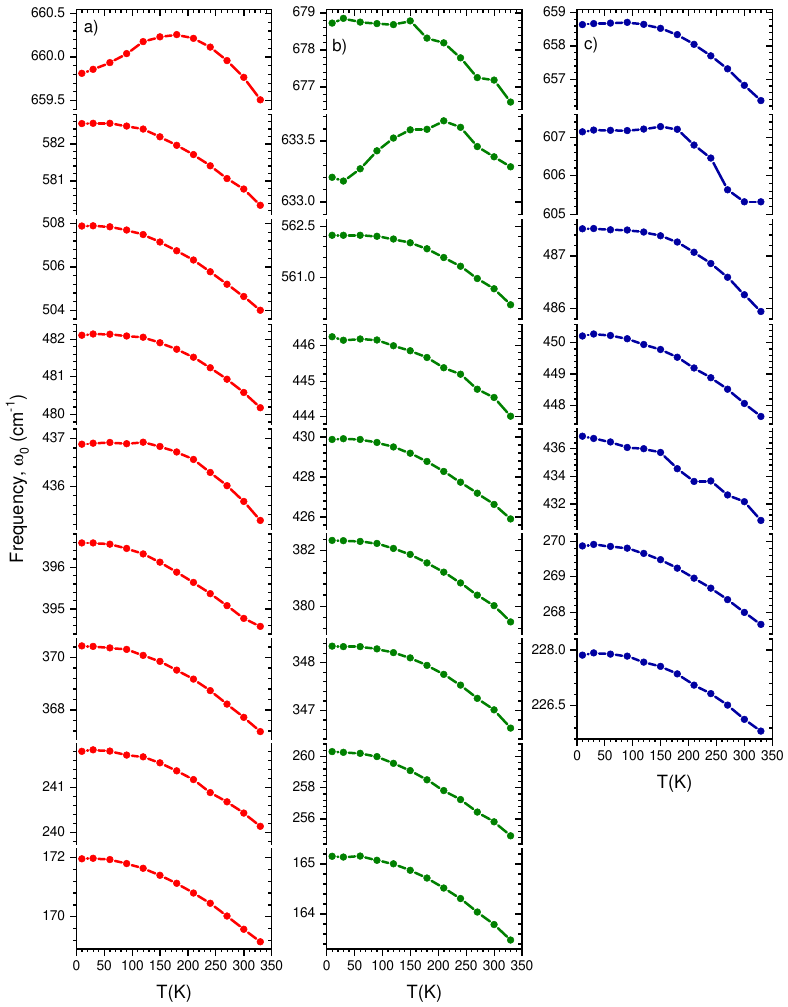}
\caption{\label{YAO temp dep omega0} Temperature dependence of the eigenfrequency $\omega_0$ of the infrared-active TO phonons of YAlO$_\text{3}$: a) $a$-axis response b) $b$-axis response c) $c$-axis response.}
\end{figure}

Finally, Fig. \ref{YAO temp dep omega0} shows the temperature dependence of the TO phonon positions between \SI{10}{\kelvin}  and \SI{330}{\kelvin}. The full set of fit parameters is available in the Supplemental Material \cite{SupMat}. The temperature dependent measurements were performed on the 3 different surface-cut samples, but in single orientation on each: the $a$-axis response was measured on the $c$-cut sample, the $b$-axis response on the $a$-cut sample and $c$-axis one on the $b$-cut sample. Once the anisotropic model was optimized on the full set of data from 6 orientations at room temperature, the reduced set of 3 orientations has been found to be sufficient to reliably model and fit the temperature dependence of the phonon parameters. Since the temperature dependent measurements have only been performed in the far-infrared part of the spectra below \SI{700}{\per\centi\meter}, the highest $\omega_{\text{LO}^*}$ feature is not included.

As there is no instability or phase transition predicted or observed for YAlO$_\text{3}$, the temperature behavior of the phonon positions is rather conventional. Only the high frequency modes, likely related to oxygen octahedra stretching, show some anomalous temperature dependence. Discussion about origin of the anomalies is beyond the scope of this work.

\section{Summary and Conclusions}
We have performed broad-band (THz to UV) spectroscopic ellipsometry experiments to determine the full dielectric tensor of the orthorhombic perovskite YAlO$_\text{3}$. This material is widely used as a substrate for thin film growth of complex oxides. The detailed knowledge of its anisotropic dielectric response function is thus a precondition for a correct analysis of the thin film response. For this reason, we cared to explain all the anisotropy-related features that appear in the raw experimental data.

The ellipsometry measurements have been performed for 6 different orientations of the plane of incidence of the photons and the sample surface with respect to the principal axes of the orthorhombic YAlO$_\text{3}$ crystal. Wedged crystals have been used to avoid artefacts that can arise from photons that are reflected from the backside surface in the frequency ranges where YAlO$_\text{3}$ is fully transparent, i.e. below about \SI{100}{\per\centi\meter} (\SI{200}{\per\centi\meter}) at room temperature (at low temperature) and well above the range of the multi-phonon excitations, i.e. at $\omega>\SI{1500}{\per\centi\meter}$. 

The measured spectra are presented in terms of the Mueller matrix elements $N$, $C$, and $S$. It is shown that their analysis with an anisotropic model provides robust and unambiguous results that capture all the essential spectroscopic features. In particular, it yields all the infrared-active phonon modes with eigenfrequencies (positions) that agree well with previous reports and theoretical predictions. We also discuss the spectral features that arise from the anisotropy of the dielectric response, including a negative refraction effect that occurs in the vicinity of the highest anisotropic LO-mode, $\omega_\text{LO}$. 

From the 6 possible high-symmetry orientations, we have compared pairs of NCS datasets taken on different crystal surfaces, but with the same crystal axis oriented along the $x$-axis of the instrument coordinate system (intersection of the plane of incidence and the sample surface). This comparison revealed a) quasi-isotropic character of the ellipsometric spectra in the ranges away from the LO frequencies – they are dominated by the response function of the crystal axis aligned with the $x$ coordinate of the ellipsometer, and b) the anisotropic effects, occurring near strongest $\omega_\text{TO}$ frequencies of the $y$-axis response, and near the $\omega_\text{LO}$ frequencies of the remaining two crystal axes, are mutually opposite when the $y$- and $z$-axis responses are interchanged.

Last but not least, we have studied the temperature dependence of the phonon parameters. Whereas no anomalous behavior has been observed, we expect that the detailed knowledge of the temperature dependent optical response of YAlO$_\text{3}$ will be useful for future studies of various thin films of complex oxides that are grown on YAlO$_\text{3}$ substrates.

\begin{acknowledgments}
This research was funded by the Swiss National Science Foundation (SNSF) through projects No. 200020-172611 and 200021-214905. Yu.G. Pashkevich acknowledges the financial support of the SNSF through the individual grants IZSEZ0 212006 and IZSEZ0 215824 of the ``Scholars at Risk'' Program. For the purpose of Open Access, a CC BY public copyright license is applied to any Author Accepted Manuscript (AAM) version arising from this submission.
\end{acknowledgments}

\bibliography{YAO_biblio.bib}

\end{document}


\title{Supplemental Material of "Broad-band ellipsometry study of the anisotropic dielectric response of YAlO$_3$"}

\author{Laurent Bugnon}
\affiliation{Department of Physics, University of Fribourg, CH-1700 Fribourg, Switzerland}
\author{Yurii G. Pashkevich}
\affiliation{Department of Physics, University of Fribourg, CH-1700 Fribourg, Switzerland}
\affiliation{Galkin Donetsk Institute for Physics and Engineering of NASU, 03028 Kyiv, Ukraine}
\author{Christian Bernhard}
\author{Premysl Marsik}
\affiliation{Department of Physics, University of Fribourg, CH-1700 Fribourg, Switzerland}

\date{\today}

\maketitle

\section{Dielectric function models}
In the main text, we have defined the factorized four-parameter model of the dielectric response functions, here omitting the $a$-, $b$-, and $c$-axis indices:

\begin{equation}
\varepsilon(\omega)=\varepsilon_{\infty} \prod_{i=1}^{N} \frac{\omega^2_{\mathrm{LO}i} - \omega^2 - i\omega\gamma_{\mathrm{LO}i}}{\omega^2_{\mathrm{TO}i} - \omega^2 - i\omega\gamma_{\mathrm{TO}i}},
\label{TOLO diel func}
\end{equation}

where the $\omega_{\text{TO}i}$, $\gamma_{\text{TO}i}$ represent the position and broadening of the TO modes, i.e. Lorentz peaks in the dielectric function $\varepsilon(\omega)$, while the $\omega_{\text{LO}i}$, $\gamma_{\text{LO}i}$ represent the position and broadening of the LO modes, Lorentzians in the inverse of the dielectric function $1/\varepsilon(\omega)$. Parameter $\varepsilon_\infty$ accounts for the high-frequency dielectric constant.

Alternatively, any of the three diagonal elements of dielectric tensor of YAlO$_\text{3}$, $\varepsilon_a$, $\varepsilon_b$ and $\varepsilon_c$, can be calculated in the infrared range with the following harmonic-oscillator approximation (HOA) as a sum of Lorentzian line-shapes corresponding to the TO modes: 
\begin{equation}
\varepsilon(\omega)=\varepsilon_{\infty} + \sum^{N}_{i=1} \frac{A_{i} \omega^2_{\mathrm{TO}i}}{\omega^2_{\mathrm{TO}i} - \omega^2 - i\omega\gamma_{\mathrm{TO}i}},
\label{HOA diel func}
\end{equation}
where the parameters $A_{i}$, $\omega_{\text{TO}i}$, $\gamma_{\text{TO}i}$ account for the dielectric strength, $A=\Delta \varepsilon(\omega=0)$, the transversal optical resonance frequency, $\omega_{\text{TO}}$, and the damping (peak broadening), $\gamma_{\text{TO}}$, of the individual oscillators. The corresponding longitudinal optical phonon frequencies, $\omega_{\text{LO}}$, can be estimated from the poles of $1/\varepsilon(\omega)$ in the limit $\gamma_{\text{TO}i} \rightarrow 0$, $\forall \, i=1 \ldots N$.

Analogically, the inverse dielectric function, $\eta(\omega)$, can be defined as:

\begin{equation}
\frac{1}{\varepsilon(\omega)} \equiv \eta(\omega)=\eta_{\infty} + \sum^{N}_{i=1} \frac{B_{i} \omega^2_{\mathrm{LO}i}}{\omega^2_{\mathrm{LO}i} - \omega^2 - i\omega\gamma_{\mathrm{LO}i}},
\label{HOA inv}
\end{equation}

with parameters $\omega_{\text{LO}i}$, $\gamma_{\text{LO}i}$ representing the positions and broadenings of the longitudinal modes. The strength parameters, $B_{i}$, must be negative, since the function $1/\varepsilon(\omega)$ has negative imaginary part. Related is the so-called Loss function, defined as negative imaginary part of the inverse dielectric function: Im\{$-1/\varepsilon(\omega)$\}, which acquires non-negative values.

Since the studied material, YAlO$_\text{3}$, does not exhibit strongly asymmetric phonon line-shapes, we were able to fit the experimental data reasonably well with all three expressions listed above. Nevertheless, the additive HOA model, eq. \eqref{HOA diel func} here, failed to reproduce the strong anisotropic features on the highest $\omega_{\text{LO}*}$  edges. This could be improved by introducing a nonzero imaginary part of $\varepsilon_\infty$, as a fit parameter, to the response functions of all thee axes. These additional parameters allowed us to obtain a good fit of the shapes of the anisotropic features at the highest $\omega_{\text{LO}^*}$. While such a procedure is somewhat unphysical, since it violates the Kramers-Kronig consistency, we found that it has a negligible impact on the spectra except for the anisotropy features at the highest $\omega_{\text{LO}^*}$.

On the other hand, the additive model for inverted dielectric function, eq. \eqref{HOA inv}, sticks better to the $\omega_{\text{LO}^*}$ related features, but does not capture well the sharp and strong TO edges in the NCS spectra.

Eventually, we used the four-parameter factorized model to precisely fit the room temperature data, as presented in the main text. The temperature dependent NCS data were analyzed with the additive HOA model, and the high energy range with the strong $\omega_{\text{LO}^*}$ features was not included in the fit (more info below in section \ref{section_tdep}).

In the table I in the main text we are also showing the dielectric strength of the TO modes as $\Delta \varepsilon(\omega=0)$. These values were obtained by comparing the factorized form, eq. \eqref{TOLO diel func}, with a corresponding additive model, eq. \eqref{HOA diel func}, for vanishing broadening parameters, $\gamma$.

Once the broadenings are omitted, the additive HOA model:

\begin{equation}
\varepsilon(\omega)=\varepsilon_{\infty} + \sum^{N}_{i=1} \frac{A_{i} \omega^2_{\mathrm{TO}i}}{\omega^2_{\mathrm{TO}i} - \omega^2},
\label{HOA wo gamma}
\end{equation}

can be factorized to

\begin{equation}
\varepsilon(\omega)=\varepsilon_{\infty} \prod_{i=1}^{N} \frac{\omega^2_{\mathrm{LO}i} - \omega^2}{\omega^2_{\mathrm{TO}i} - \omega^2},
\label{FPSQ wo gamma}
\end{equation}

if the strength parameters, $A=\Delta \varepsilon$, satisfy:

\begin{equation}
A_i=\frac{\varepsilon_{\infty}}{\omega^2_{\mathrm{TO}i}}
\frac{ \prod_{k} \omega^2_{\mathrm{LO}k} - \omega^2_{\mathrm{TO}i} }
{\prod_{k\neq i} \omega^2_{\mathrm{TO}k} - \omega^2_{\mathrm{TO}i} }.
\label{delta_eps}
\end{equation}

The $\Delta \varepsilon$ values listed in table I of the main text were calculated from the $\omega_{\text{TO}}$ and $\omega_{\text{LO}}$ using the above equation.

\newpage
\section{Room temperature NCS data} 

Here we present the same YAlO$_\text{3}$ room temperature far-infrared NCS data and models as in the main text (figures 2a, 3a, 4a), but in an alternative arrangement, showing together the two spectra that can be obtained for a single sample with a high symmetry surface cut and in two high symmetry orientations.

Additionally, on the three following figures, we show all the $\omega_{\text{TO}}$ and $\omega_{\text{LO}}$ frequencies, as listed in the table I in the main text, with vertical solid and dashed lines, respectively.

\begin{figure}[!htb]
\includegraphics[width=\textwidth]{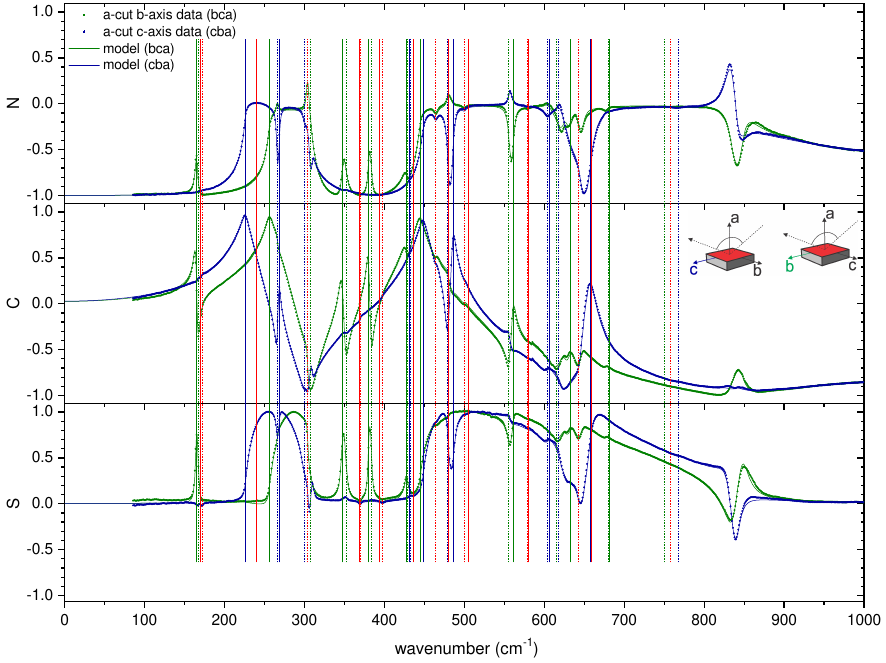}
\caption{\label{YAO a-cut RT} Far-infrared NCS ellipsometry spectra of YAlO$_\text{3}$ at \SI{300}{\kelvin} with angle of incidence $\varphi = \SI{75}{\degree}$ measured on $a$-cut sample, with the $b$-axis (green points) and $c$-axis (blue points) parallel to the plane of incidence. Sketches of the measurement configurations are shown in the inset of the middle panel. Also shown are the best fits with the anisotropic model (green and blue lines). Vertical lines mark the $\omega_{\text{TO}}$ (solid lines) and $\omega_{\text{LO}}$ (dashed lines) frequencies of the infrared active phonons in the response functions of the $a$-axis (red colour), $b$-axis (green colour), and $c$-axis (blue colour).}
\end{figure}

\begin{figure}[!htb]
\includegraphics[width=\textwidth]{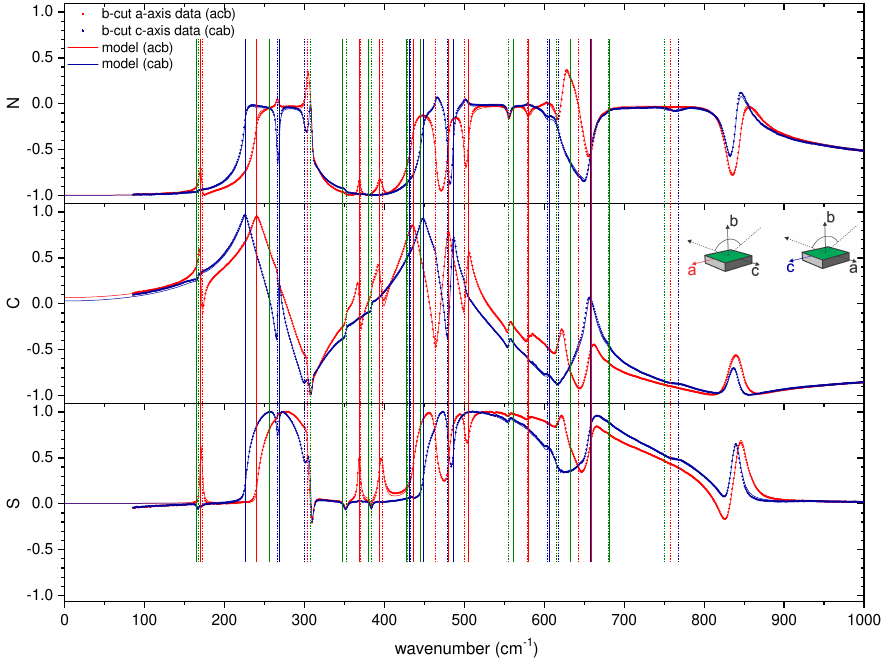}
\caption{\label{YAO b-cut RT}  Far-infrared NCS ellipsometry spectra of YAlO$_\text{3}$ at \SI{300}{\kelvin} with angle of incidence $\varphi = \SI{75}{\degree}$ measured on $b$-cut sample, with the $a$-axis (red points) and $c$-axis (blue points) parallel to the plane of incidence. Sketches of the measurement configurations are shown in the inset of the middle panel. Also shown are the best fits with the anisotropic model (red and blue lines). Vertical lines mark the $\omega_{\text{TO}}$ (solid lines) and $\omega_{\text{LO}}$ (dashed lines) frequencies of the infrared active phonons in the response functions of the $a$-axis (red colour), $b$-axis (green colour), and $c$-axis (blue colour).}
\end{figure}

\begin{figure}[!htb]
\includegraphics[width=\textwidth]{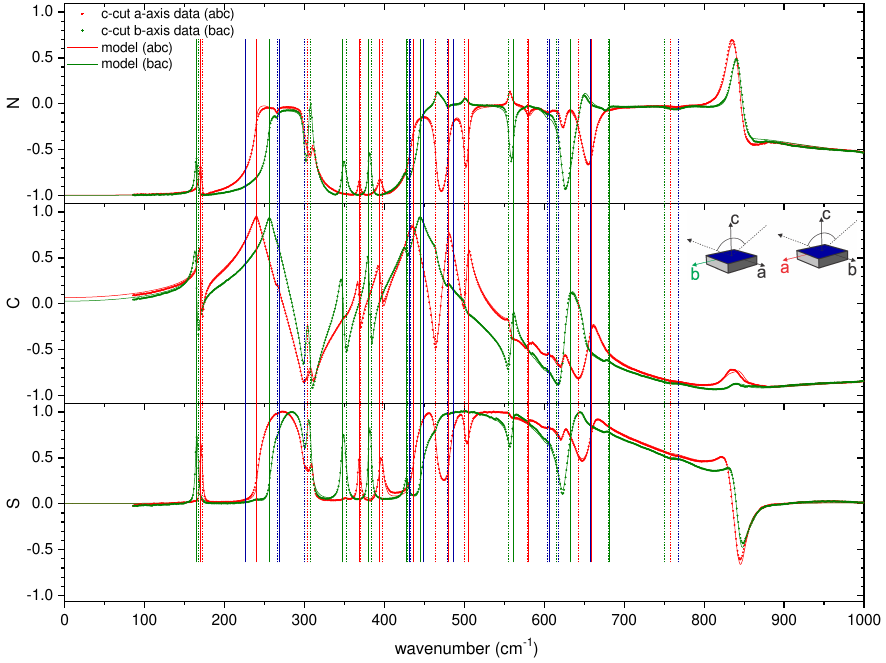}
\caption{\label{YAO c-cut RT} Far-infrared NCS ellipsometry spectra of YAlO$_\text{3}$ at \SI{300}{\kelvin} with angle of incidence $\varphi = \SI{75}{\degree}$ measured on $c$-cut sample, with the $a$-axis (red points) and $b$-axis (green points) parallel to the plane of incidence. Sketches of the measurement configurations are shown in the inset of the middle panel. Also shown are the best fits with the anisotropic model (red and green lines). Vertical lines mark the $\omega_{\text{TO}}$ (solid lines) and $\omega_{\text{LO}}$ (dashed lines) frequencies of the infrared active phonons in the response functions of the $a$-axis (red colour), $b$-axis (green colour), and $c$-axis (blue colour).}
\end{figure}

\section{Model dielectric response of YAO} 

Here we show the same model of the dielectric response of YAlO$_\text{3}$ at room temperature as discussed in the main text and plotted on figure 5 there. For better clarity, here we plot the real and imaginary parts of the model dielectric functions for the 3 diagonal elements of the permittivity tensor on separate panels and also show the inverted dielectric function  $1/\varepsilon(\omega)$.

Additionally, we show the results of an unconstrained point-by-point fit, that is, at each spectral point independently, we fit the values of the three complex response functions, $\varepsilon_a$, $\varepsilon_b$ and $\varepsilon_c$ (6 real parameters), simultaneaously to the full set of NCS data obtained from the 6 high symmetry configurations (18 datapoints). At each spectral point the fit is started from the already optimized factorized model.

Both models, parameterized and point-by-point, were fitted to the same dataset merged from the far-infrared RAE data and the mid-infrared RCE data. The two segments were merged at \SI{540}{\per\centi\meter}. Due to imperfect overlap between the two segments, there is a noticeable glitch in the point-by-point data at the merge point.

\begin{figure}[!htb]
\includegraphics[width=\textwidth]{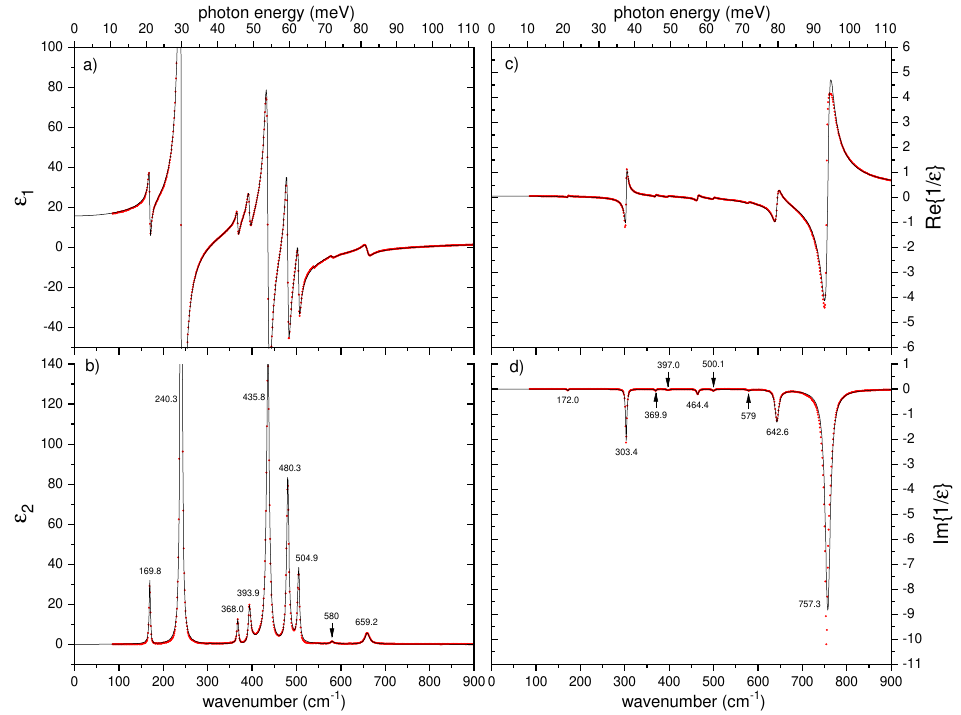}
\caption{\label{YAOa_epsinvpbp}  Far-infrared response along the $a$-axis of the orthorhombic YAlO$_\text{3}$ dielectric tensor, $\varepsilon_a(\omega)$, resulting from analysis with four-parameter factorized model (black lines) and unconstrained point-by-point fit (red symbols), shown as: a) real part of the dielectric function, $\varepsilon_1(\omega)$; b) imaginary part of the dielectric function, $\varepsilon_2(\omega)$; c) real part of the inverse dielectric function, Re\{$1/\varepsilon(\omega)$\}; d) imaginary part of the inverse dielectric function Im\{$1/\varepsilon(\omega)$\}. The labels on panels b) and d) mark the positions of the $\omega_{\text{TO}}$ and $\omega_{\text{LO}}$ frequencies (in \SI{}{\per\centi\meter}), respectively.}
\end{figure}

\begin{figure}[!htb]
\includegraphics[width=\textwidth]{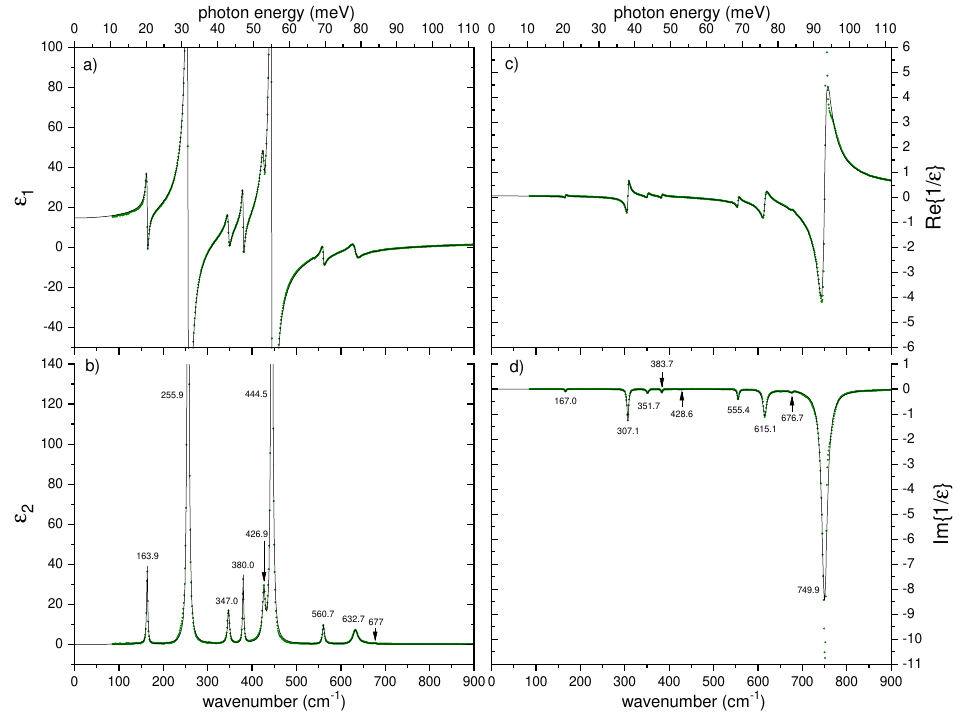}
\caption{\label{YAOb_epsinvpbp}  Far-infrared response along the $b$-axis of the orthorhombic YAlO$_\text{3}$ dielectric tensor, $\varepsilon_b(\omega)$, resulting from analysis with four-parameter factorized model (black lines) and unconstrained point-by-point fit (green symbols), shown as: a) real part of the dielectric function, $\varepsilon_1(\omega)$; b) imaginary part of the dielectric function, $\varepsilon_2(\omega)$; c) real part of the inverse dielectric function, Re\{$1/\varepsilon(\omega)$\}; d) imaginary part of the inverse dielectric function Im\{$1/\varepsilon(\omega)$\}. The labels on panels b) and d) mark the positions of the $\omega_{\text{TO}}$ and $\omega_{\text{LO}}$ frequencies (in \SI{}{\per\centi\meter}), respectively.}
\end{figure}

\begin{figure}[!htb]
\includegraphics[width=\textwidth]{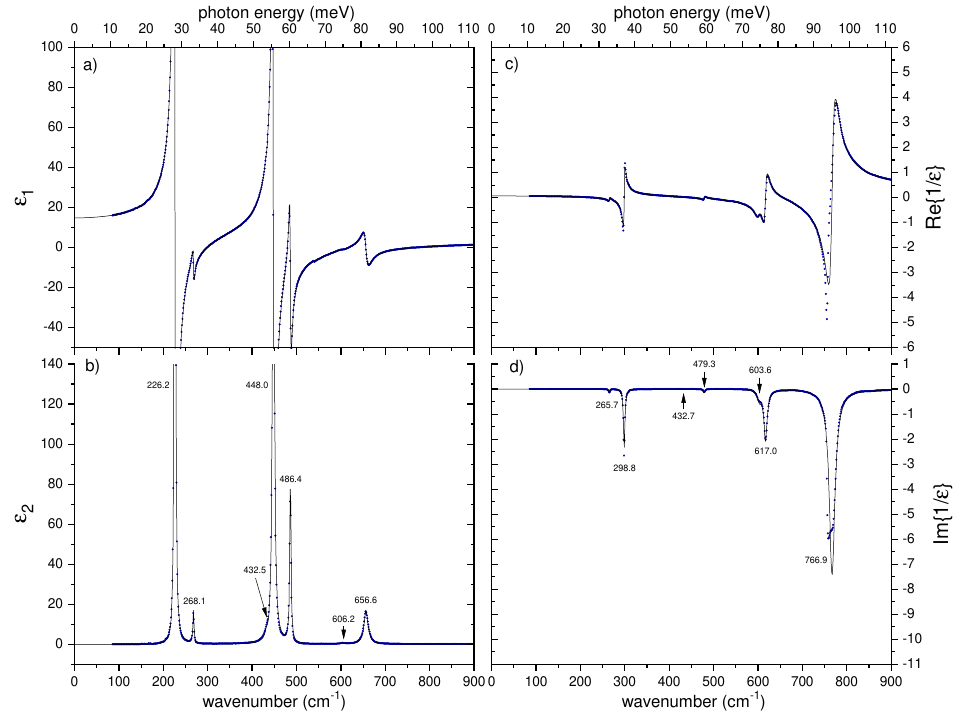}
\caption{\label{YAOc_epsinvpbp}  Far-infrared response along the $c$-axis of the orthorhombic YAlO$_\text{3}$ dielectric tensor, $\varepsilon_c(\omega)$, resulting from analysis with four-parameter factorized model (black lines) and unconstrained point-by-point fit (green symbols), shown as: a) real part of the dielectric function, $\varepsilon_1(\omega)$; b) imaginary part of the dielectric function, $\varepsilon_2(\omega)$; c) real part of the inverse dielectric function, Re\{$1/\varepsilon(\omega)$\}; d) imaginary part of the inverse dielectric function Im\{$1/\varepsilon(\omega)$\}. The labels on panels b) and d) mark the positions of the $\omega_{\text{TO}}$ and $\omega_{\text{LO}}$ frequencies (in \SI{}{\per\centi\meter}), respectively.}
\end{figure}

\section{Low temperature NCS simulated} 

At low temperatures, the phonon line-shapes become very narrow, which makes the features observed in the experimental NCS data more pronounced and distinct. To better illustrate the anisotropy effects that were discussed in the main text, we show here comparable figures of the NCS spectra for all orientations at \SI{10}{\kelvin}. Vertical lines represent the TO and LO positions as obtained from the factorized model fitted to the \SI{10}{\kelvin} data. The next three figures show only the model lines for greater clarity, while the comparisons of data and model are shown in the next section, further below.

\begin{figure}[!htb]
\includegraphics[width=\textwidth]{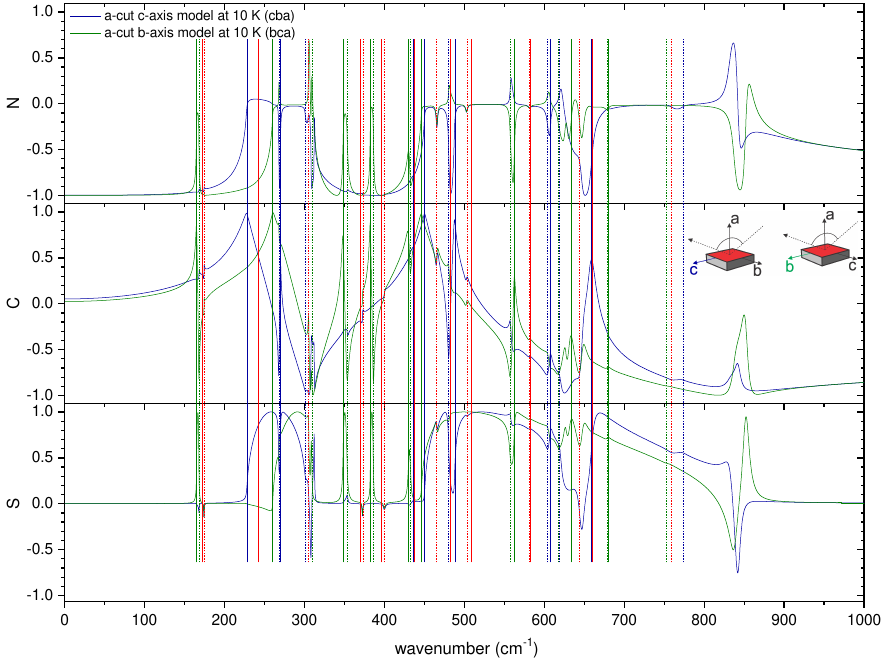}
\caption{\label{YAO a-cut LT} Simulated far-infrared NCS ellipsometry spectra of YAlO$_\text{3}$ at \SI{10}{\kelvin} with angle of incidence $\varphi = \SI{75}{\degree}$ for $a$-cut sample, with the $b$-axis (green lines) and $c$-axis (blue lines) parallel to the plane of incidence. Sketches of the simulated configurations are shown in the inset of the middle panel. Simulation was calculated using model parameters fitted to \SI{10}{\kelvin} experimental data. Vertical lines mark the $\omega_{\text{TO}}$ (solid lines) and $\omega_{\text{LO}}$ (dashed lines) frequencies of the infrared active phonons in the response functions of the $a$-axis (red colour), $b$-axis (green colour), and $c$-axis (blue colour), at \SI{10}{\kelvin}. Note that $c$-axis response has a weak phonon at \SI{436.3}{\per\centi\meter} with TO-LO splitting of \SI{0.15}{\per\centi\meter}, which renders the pair of vertical lines indistinguishable.}
\end{figure}

\begin{figure}[!htb]
\includegraphics[width=\textwidth]{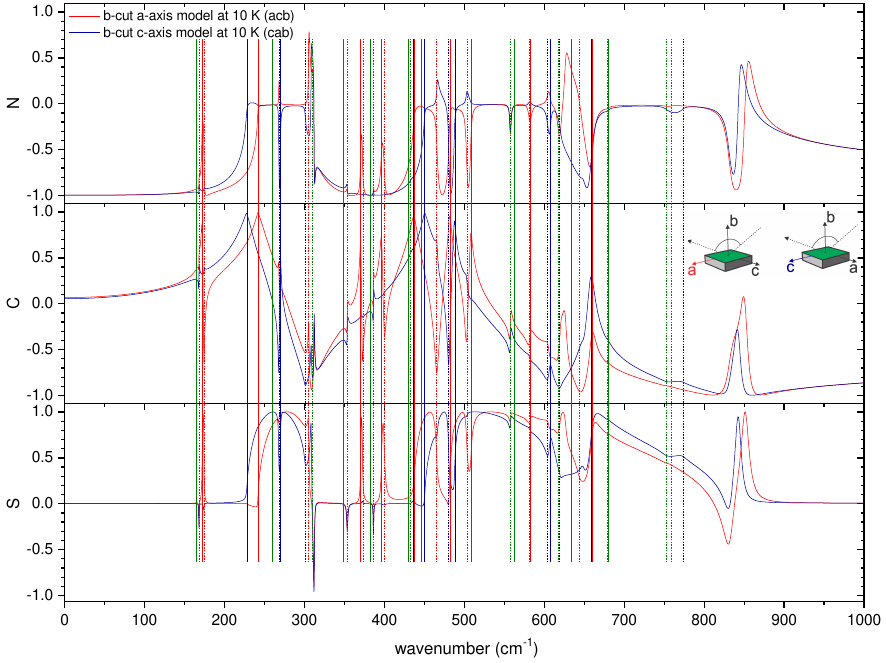}
\caption{\label{YAO b-cut LT} Simulated far-infrared NCS ellipsometry spectra of YAlO$_\text{3}$ at \SI{10}{\kelvin} with angle of incidence $\varphi = \SI{75}{\degree}$ for $b$-cut sample, with the $a$-axis (red lines) and $c$-axis (blue lines) parallel to the plane of incidence. Sketches of the simulated configurations are shown in the inset of the middle panel. Simulation was calculated using model parameters fitted to \SI{10}{\kelvin} experimental data. Vertical lines mark the $\omega_{\text{TO}}$ (solid lines) and $\omega_{\text{LO}}$ (dashed lines) frequencies of the infrared active phonons in the response functions of the $a$-axis (red colour), $b$-axis (green colour), and $c$-axis (blue colour), at \SI{10}{\kelvin}. Note that $c$-axis response has a weak phonon at \SI{436.3}{\per\centi\meter} with TO-LO splitting of \SI{0.15}{\per\centi\meter}, which renders the pair of vertical lines indistinguishable.}
\end{figure}

\begin{figure}[!htb]
\includegraphics[width=\textwidth]{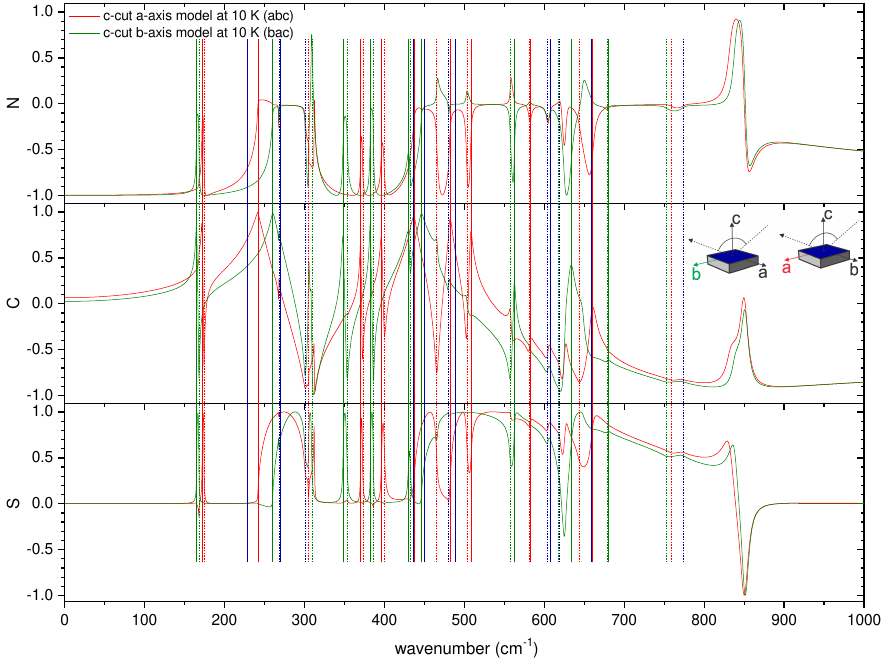}
\caption{\label{YAO c-cut LT} Simulated far-infrared NCS ellipsometry spectra of YAlO$_\text{3}$ at \SI{10}{\kelvin} with angle of incidence $\varphi = \SI{75}{\degree}$ for $c$-cut sample, with the $a$-axis (red lines) and $b$-axis (green lines) parallel to the plane of incidence. Sketches of the simulated configurations are shown in the inset of the middle panel. Simulation was calculated using model parameters fitted to \SI{10}{\kelvin} experimental data. Vertical lines mark the $\omega_{\text{TO}}$ (solid lines) and $\omega_{\text{LO}}$ (dashed lines) frequencies of the infrared active phonons in the response functions of the $a$-axis (red colour), $b$-axis (green colour), and $c$-axis (blue colour), at \SI{10}{\kelvin}. Note that $c$-axis response has a weak phonon at \SI{436.3}{\per\centi\meter} with TO-LO splitting of \SI{0.15}{\per\centi\meter}, which renders the pair of vertical lines indistinguishable.}
\end{figure}

\section{Low temperature data and model compared} 

Comparison of NCS data of YAlO$_\text{3}$, obtained at \SI{10}{\kelvin}, and the best-fit four-parameter factorized model. The low temperature measurements were done only on three of the six possible high-symmetry configurations, and only with the RAE far-infrared setup. Following three figures show the raw data that contain several bands of low/no signal. Noisy band between \SIrange{700}{750}{\per\centi\meter} is caused by the absorption in Polyethylene, used as window of the far-infrared bolometer and substrate of the polarizers. Band between \SIrange{600}{620}{\per\centi\meter}, visible in the $S$-element spectrum (bottom panel), is caused by the absorption in the Silicon retarder. Feature above \SI{900}{\per\centi\meter} is an artefact and should be ignored.

\begin{figure}[!htb]
\includegraphics[width=\textwidth]{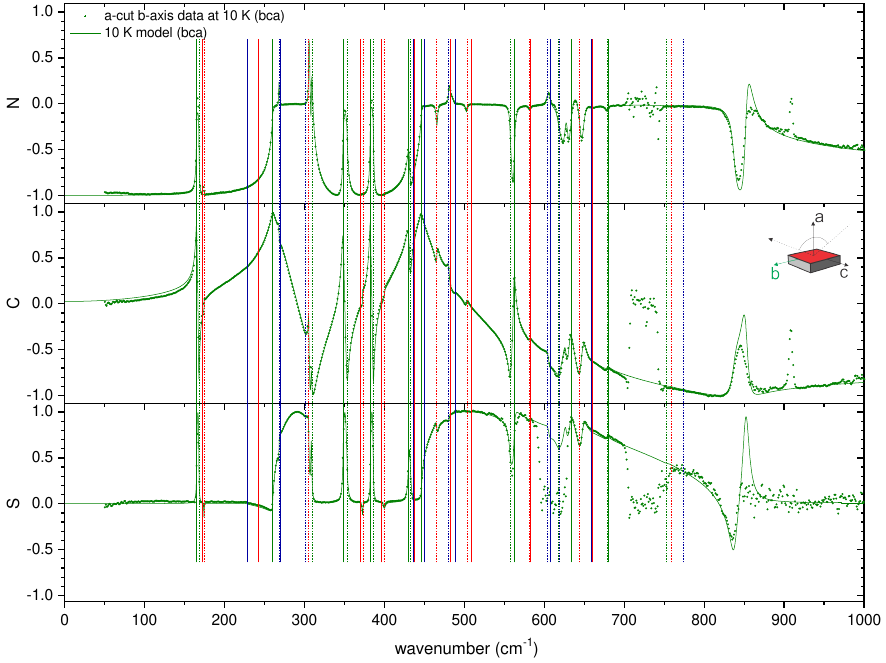}
\caption{\label{YAO a-cut data LT}  Far-infrared NCS ellipsometry spectra of YAlO$_\text{3}$ at \SI{10}{\kelvin} with angle of incidence $\varphi = \SI{75}{\degree}$ measured on $a$-cut sample, with the $b$-axis parallel to the plane of incidence (green points). Sketch of the measurement configuration is shown in the inset of the middle panel. Also shown are the best fits with the anisotropic model (green lines). Vertical lines mark the $\omega_{\text{TO}}$ (solid lines) and $\omega_{\text{LO}}$ (dashed lines) frequencies of the infrared active phonons in the response functions of the $a$-axis (red colour), $b$-axis (green colour), and $c$-axis (blue colour), at \SI{10}{\kelvin}. Note that $c$-axis response has a weak phonon at \SI{436.3}{\per\centi\meter} with TO-LO splitting of \SI{0.15}{\per\centi\meter}, which renders the pair of vertical lines indistinguishable.}
\end{figure}

\begin{figure}[!htb]
\includegraphics[width=\textwidth]{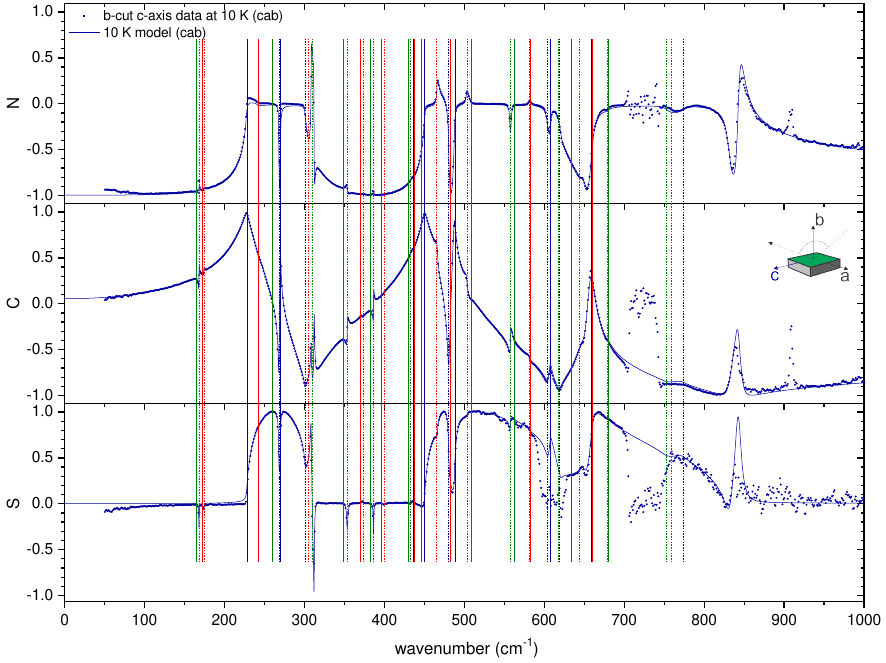}
\caption{\label{YAO b-cut data LT} Far-infrared NCS ellipsometry spectra of YAlO$_\text{3}$ at \SI{10}{\kelvin} with angle of incidence $\varphi = \SI{75}{\degree}$ measured on $b$-cut sample, with the $c$-axis parallel to the plane of incidence (blue points). Sketch of the measurement configuration is shown in the inset of the middle panel. Also shown are the best fits with the anisotropic model (blue lines). Vertical lines mark the $\omega_{\text{TO}}$ (solid lines) and $\omega_{\text{LO}}$ (dashed lines) frequencies of the infrared active phonons in the response functions of the $a$-axis (red colour), $b$-axis (green colour), and $c$-axis (blue colour), at \SI{10}{\kelvin}. Note that $c$-axis response has a weak phonon at \SI{436.3}{\per\centi\meter} with TO-LO splitting of \SI{0.15}{\per\centi\meter}, which renders the pair of vertical lines indistinguishable.}
\end{figure}

\begin{figure}[!htb]
\includegraphics[width=\textwidth]{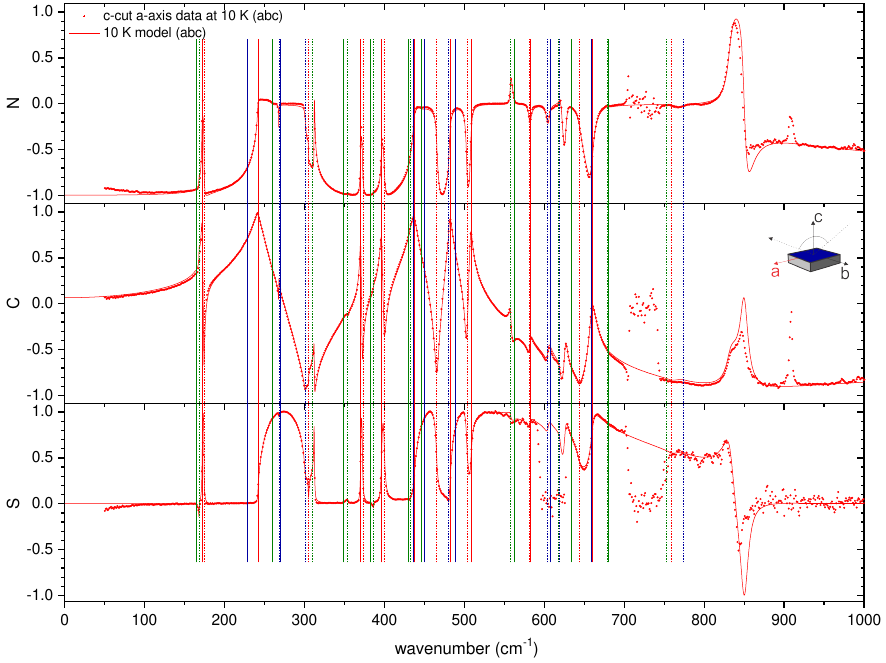}
\caption{\label{YAO c-cut data LT} Far-infrared NCS ellipsometry spectra of YAlO$_\text{3}$ at \SI{10}{\kelvin} with angle of incidence $\varphi = \SI{75}{\degree}$ measured on $c$-cut sample, with the $a$-axis parallel to the plane of incidence (red points). Sketch of the measurement configuration is shown in the inset of the middle panel. Also shown are the best fits with the anisotropic model (red lines). Vertical lines mark the $\omega_{\text{TO}}$ (solid lines) and $\omega_{\text{LO}}$ (dashed lines) frequencies of the infrared active phonons in the response functions of the $a$-axis (red colour), $b$-axis (green colour), and $c$-axis (blue colour), at \SI{10}{\kelvin}. Note that $c$-axis response has a weak phonon at \SI{436.3}{\per\centi\meter} with TO-LO splitting of \SI{0.15}{\per\centi\meter}, which renders the pair of vertical lines indistinguishable.}
\end{figure}

\section{Temperature dependent data and model parameters} \label{section_tdep}
In this section we show the temperature evolution of the far-infrared NCS experimental spectra of YAlO$_\text{3}$. As discussed in the previous section, the data originate from the far-infrared RAE setup and contain several bands of poor signal to noise. These bands were erased from the data shown on the following three figures.

Finally, temperature dependence of the model parameters is shown further below. The temperature-dependent spectra were fitted in the range \SIrange{150}{700}{\per\centi\meter} with the three-parameter additive HOA model.

\begin{figure}[!htb]
\includegraphics[width=\textwidth]{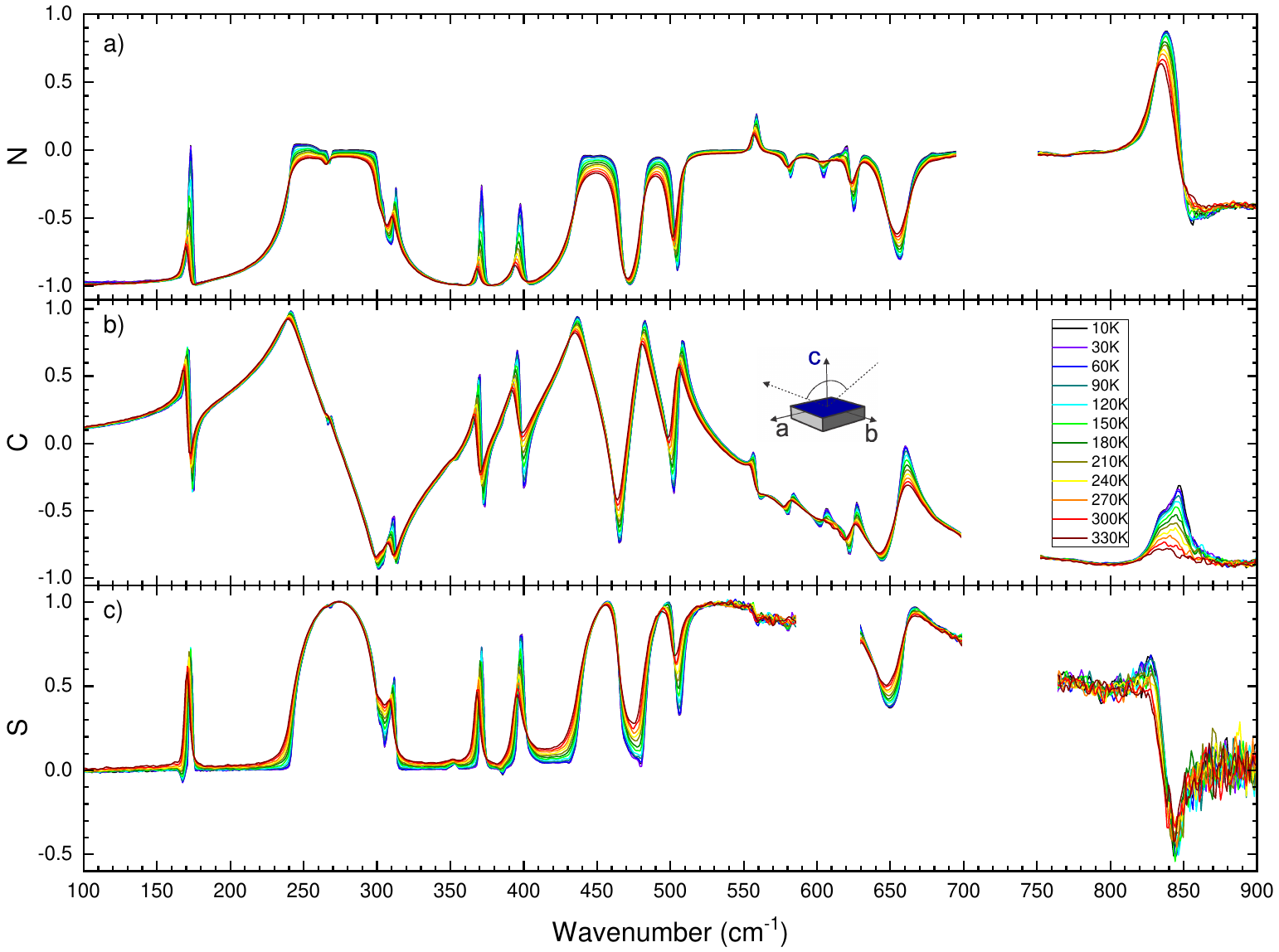}
\caption{\label{YAO c-cut a-axis}  Far-infrared response along the $a$-axis measured on $c$-cut YAlO$_\text{3}$ at angle of incidence $\varphi = \SI{75}{\degree}$, at temperatures ranging from \SIrange{10}{330}{\kelvin}. Panels a), b) and c) show the $N$, $C$ and $S$ element of the Mueller Matrix, respectively.}
\end{figure}

\begin{figure}[!htb]
\includegraphics[width=\textwidth]{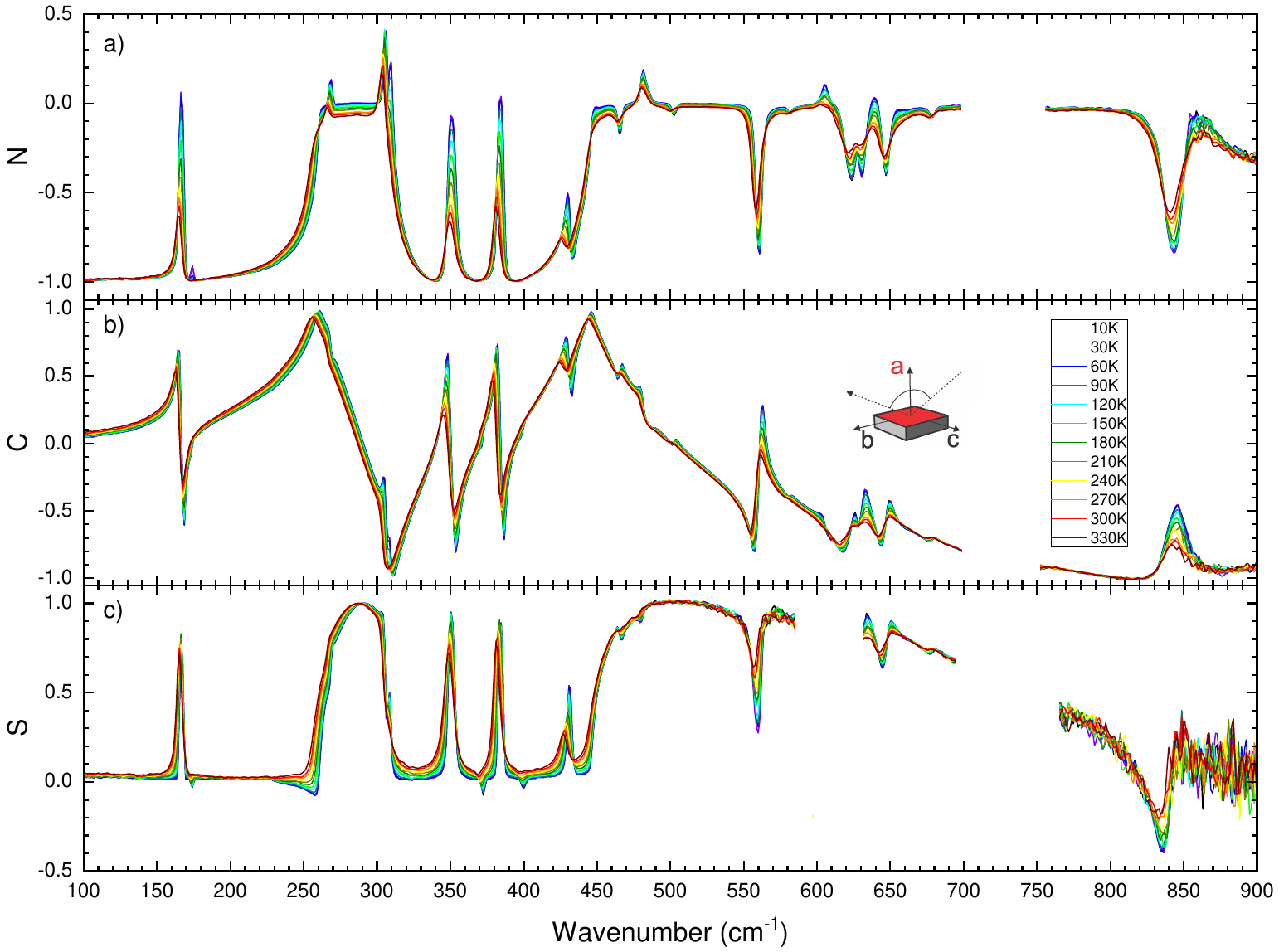}
\caption{\label{YAO a-cut b-axis} Far-infrared response along the $b$-axis measured on $a$-cut YAlO$_\text{3}$ at angle of incidence $\varphi = \SI{75}{\degree}$, at temperatures ranging from \SIrange{10}{330}{\kelvin}. Panels a), b) and c) show the $N$, $C$ and $S$ element of the Mueller Matrix, respectively.}
\end{figure}

\begin{figure}[!htb]
\includegraphics[width=\textwidth]{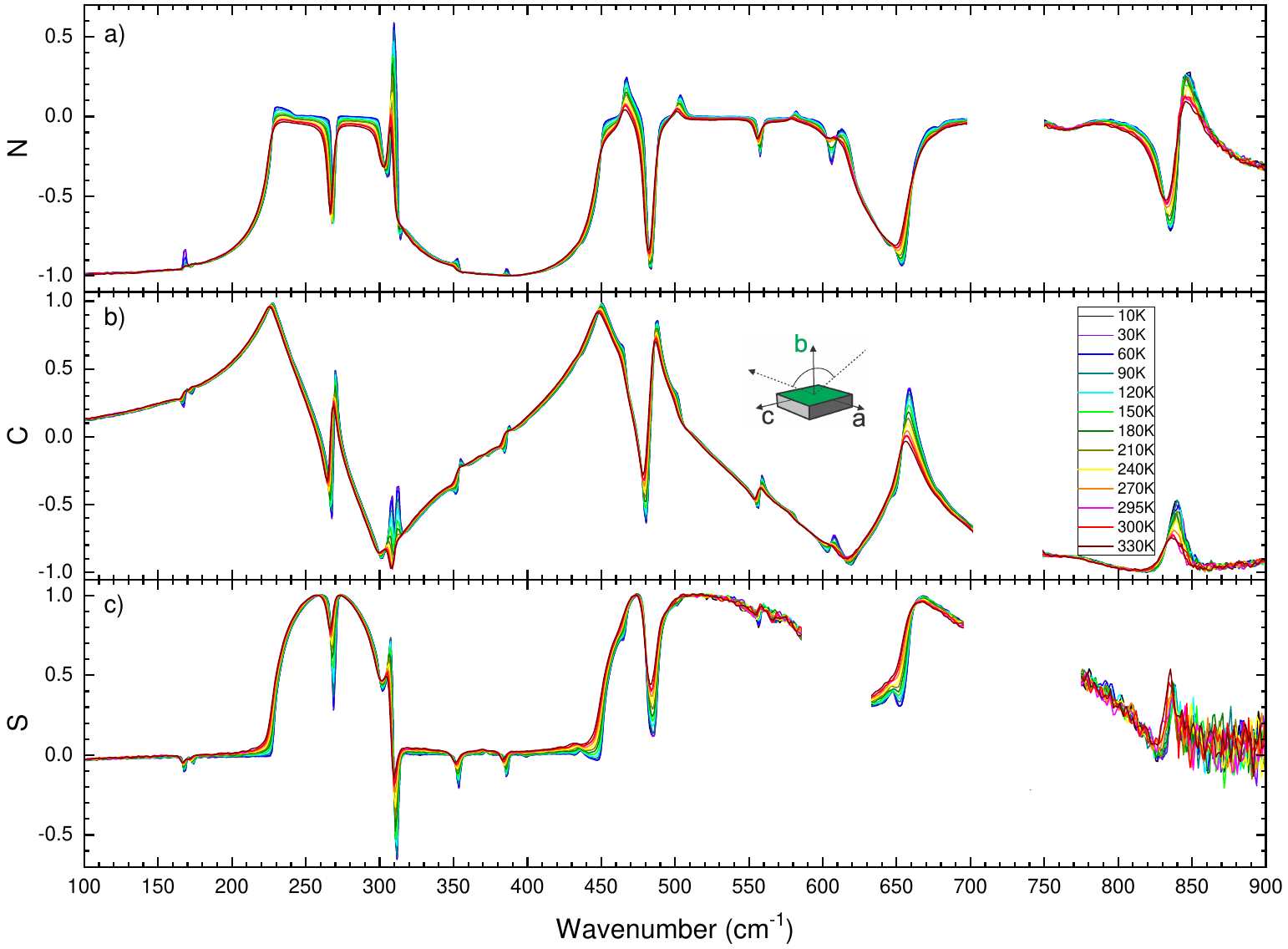}
\caption{\label{YAO b-cut c-axis}  Far-infrared response along the $c$-axis measured on $b$-cut YAlO$_\text{3}$ at angle of incidence $\varphi = \SI{75}{\degree}$, at temperatures ranging from \SIrange{10}{330}{\kelvin}. Panels a), b) and c) show the $N$, $C$ and $S$ element of the Mueller Matrix, respectively.}
\end{figure}

\begin{figure}[!htb]
\includegraphics[width=\textwidth]{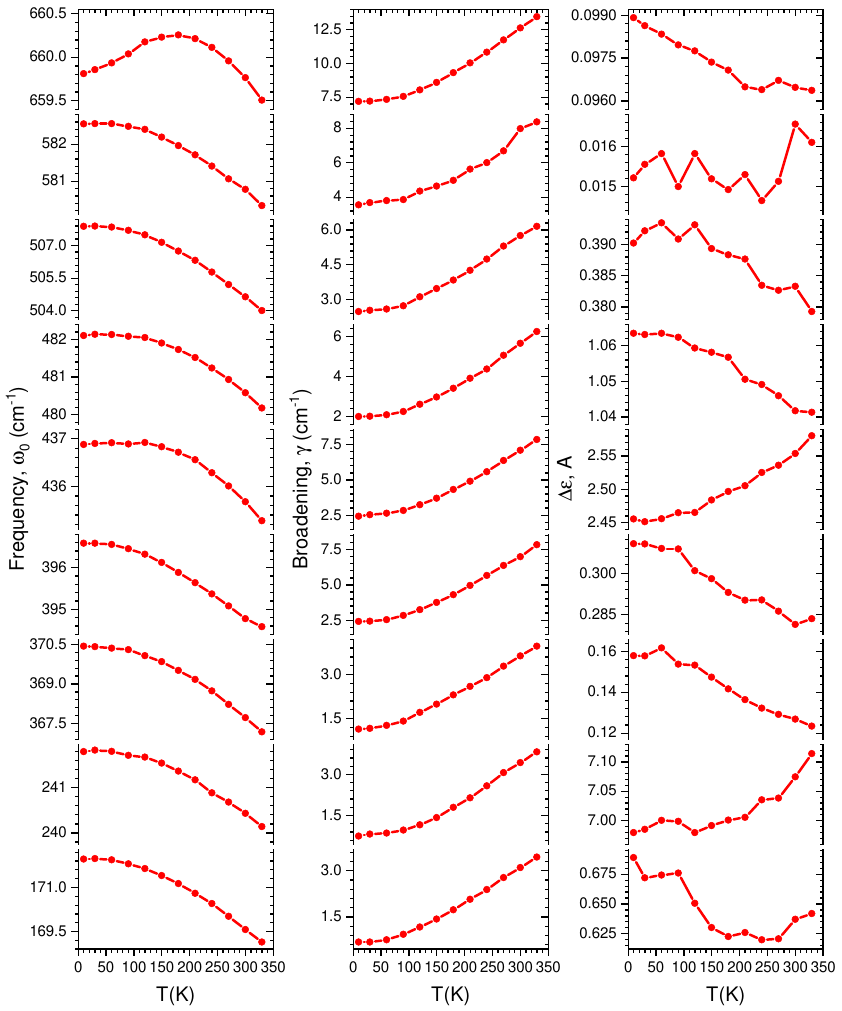}
\caption{\label{YAO c-cut a-axis parameters} Temperature dependence of YAlO$_\text{3}$ $a$-axis infrared-active TO phonons ($B_\text{1u}$) in terms of eigenfrequency $\omega_0$, broadening $\gamma$ and strength $A=\Delta\varepsilon$, at temperatures from \SIrange{10}{330}{\kelvin}.}
\end{figure}

\begin{figure}[!htb]
\includegraphics[width=\textwidth]{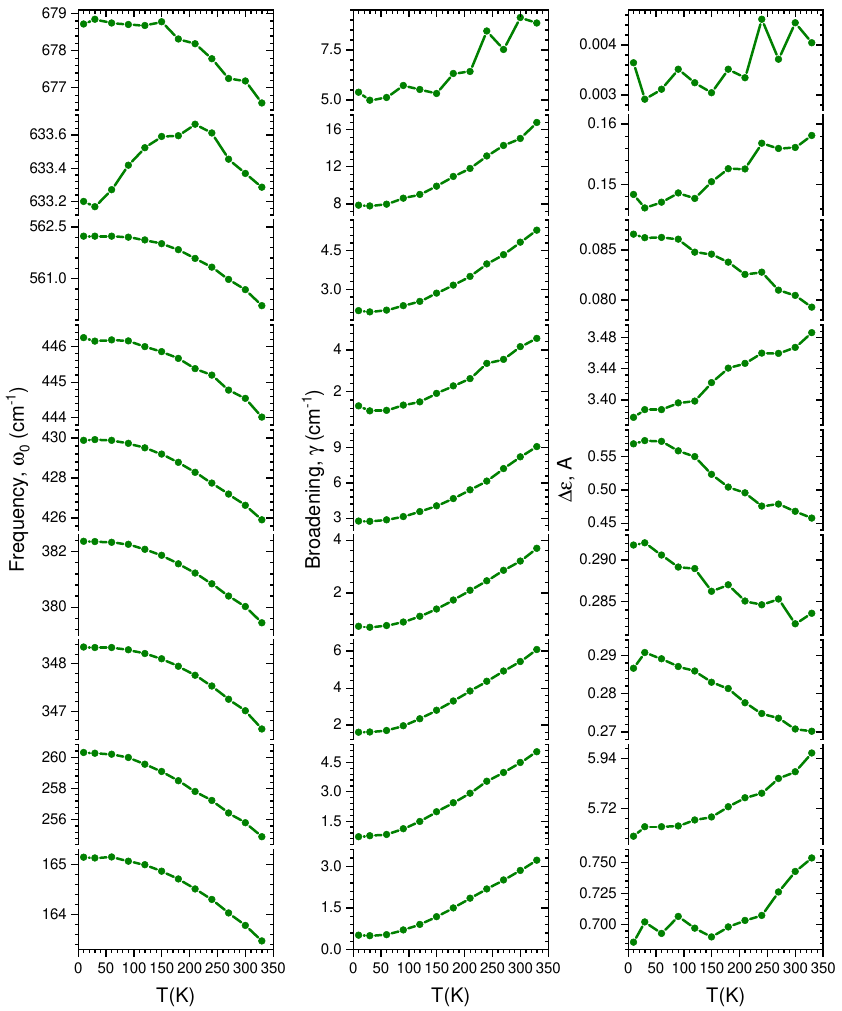}
\caption{\label{YAO a-cut b-axis parameters} Temperature dependence of YAlO$_\text{3}$ $b$-axis infrared-active TO phonons ($B_\text{3u}$) in terms of eigenfrequency $\omega_0$, broadening $\gamma$ and strength $A=\Delta\varepsilon$, at temperatures from \SIrange{10}{330}{\kelvin}.}
\end{figure}

\begin{figure}[!htb]
\includegraphics[width=\textwidth]{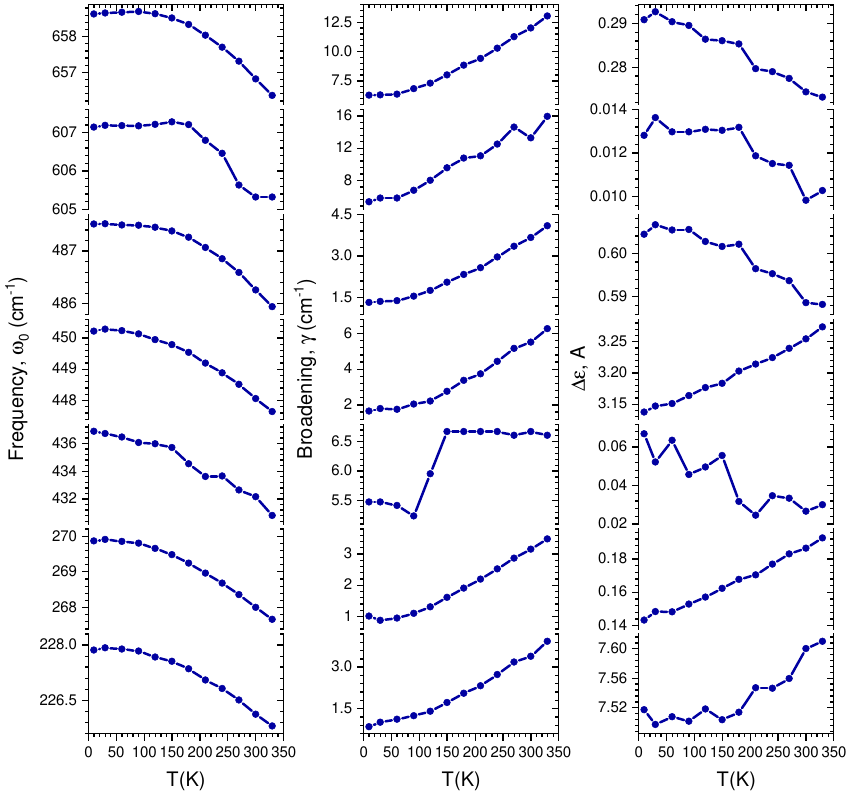}
\caption{\label{YAO b-cut c-axis parameters} Temperature dependence of YAlO$_\text{3}$ $c$-axis infrared-active TO phonons ($B_\text{2u}$) in terms of eigenfrequency $\omega_0$, broadening $\gamma$ and strength $A=\Delta\varepsilon$, at temperatures from \SIrange{10}{330}{\kelvin}.}
\end{figure}